\documentclass[11pt]{article}
\usepackage{epsfig} 
\usepackage{graphics}
\newcommand{\sect}[1]{ \section{#1} \setcounter{equation}{0} }

\newcommand{\pslash}{p \! \! \! /}

\newcommand{\xslash}{x \! \! \! /}
\newcommand{\partialslash}{\partial \! \! \! /}
\newcommand{\half}{\mbox{\small{$\frac{1}{2}$}}}

\newcommand{\threehalves}{\mbox{\small{$\frac{3}{2}$}}}

\newcommand{\Nf}{N_{\!f}}

\begin{document}

\title{Large $N$ critical exponents for the chiral Heisenberg Gross-Neveu
universality class} 
\author{J.A. Gracey, \\ Theoretical Physics Division, \\ 
Department of Mathematical Sciences, \\ University of Liverpool, \\ P.O. Box 
147, \\ Liverpool, \\ L69 3BX, \\ United Kingdom.} 
\date{}
\maketitle 

\vspace{5cm} 
\noindent 
{\bf Abstract.} We compute the large $N$ critical exponents $\eta$, $\eta_\phi$
and $1/\nu$ in $d$-dimensions in the chiral Heisenberg Gross-Neveu model to
several orders in powers of $1/N$. For instance, the large $N$ conformal
bootstrap method is used to determine $\eta$ at $O(1/N^3)$ while the other
exponents are computed to $O(1/N^2)$. Estimates of the exponents for a phase
transition in graphene are given which are shown to be commensurate with other
approaches. In particular the behaviour of the exponents in $2$~$<$~$d$~$<$~$4$
is in qualitative agreement with a functional renormalization group analysis. 
The $\epsilon$-expansion of each exponent near four dimensions is in exact 
agreement with recent four loop perturbation theory.

\vspace{-18cm}
\hspace{13.5cm}
{\bf LTH 1147}

\newpage 

\sect{Introduction.}

One of the more remarkable fundamental quantum field theories is the 
Gross-Neveu or Ashkin-Teller model, \cite{1,2}. It is a purely fermionic theory
with a sole quartic self-interaction. In some sense it is the parallel to the
more widely studied purely scalar quartic field theory which is renormalizable
in four spacetime dimensions. The spontaneous symmetry breaking phase in that
model is the basis for the Higgs mechanism of the Standard Model. By contrast
the $O(N)$ Gross-Neveu model is asymptotically free, \cite{1}, but appears to 
have little physical application since it is only renormalizable in two 
dimensions. Put another way prior to the discovery of the $W$ and $Z$ vector 
bosons of the Standard Model the physics of the weak interactions was described
by an effective field theory involving $4$-point fermion interactions. The
restriction to being effective meant that such interactions could only be
reliable as a model of Nature up to a specific momentum scale. However, the 
general Gross-Neveu class of quantum field theories, several of which were
introduced in \cite{1}, have enjoyed a renaissance over recent years. This is 
in the main due to the fact that certain phase transitions in graphene can be 
described by specific universality classes based on the Gross-Neveu model,
\cite{1}, where the classes derive from the underlying symmetry of the 
transition. For instance, stretching a graphene sheet can bring about a 
transition from a conductor to a Mott-insulating phase, \cite{3,4}. It has been
suggested that the physics of this transition can be described by what is 
termed the chiral Heisenberg Gross-Neveu model, \cite{5,6,7,8,9}. Therefore 
there has been activity in computing and estimating the fundamental critical 
exponents of the transition in this model. Prior to $2014$ there were virtually
no deep computations where the theoretical status can be appreciated in the 
right panels of Figures $1$, $2$ and $3$ given in \cite{9}. 

More specifically in \cite{9} the critical exponents $\eta$, $\eta_\phi$ and 
$1/\nu$ were estimated in spacetime dimension $d$ where $2$~$<$~$d$~$<$~$4$ 
using functional renormalization group techniques. Hence values were given for 
the three dimensional case of physical interest to graphene. By contrast at 
that time only two loop $\epsilon$ expansion results from four dimensions were 
available to compare with, \cite{10,11}. Since then this four dimensional 
perturbative work has been extended to three and four loops respectively in 
\cite{12,13,1}. The situation can be compared to the more widely studied Ising 
Gross-Neveu model which was given in the left panels of Figures $1$, $2$ and 
$3$ of \cite{9}. By contrast with the chiral Heisenberg Gross-Neveu model in 
the Ising case there are now two dimensional $\epsilon$ expansion estimates to 
four loops, \cite{15}, as well as four loop four dimensional perturbative 
information, \cite{11}. These higher loop results postdate the lower loop 
results of \cite{1,16,17,18,19,20,21} which were the state of the art at the 
time of \cite{9} for the Ising Gross-Neveu universality class. On top of this, 
Monte Carlo data, \cite{11}, as well as estimates from several orders in the 
large $N$ expansion, \cite{22,23,24,25,26,27,28}, were available to give 
independent analyses of the three dimensional exponents. Overall there was 
solid agreement for the exponents $\eta_\phi$ and $1/\nu$ from all these 
methods as well as the functional renormalization group analysis itself which 
was given in \cite{9}. However those from the $\epsilon$ expansions near two 
and four dimensions for $\eta$ were not in close agreement with the other 
methods, \cite{9}. One simple reason for this is clear from the plot for $\eta$
in Figure $1$ of \cite{9}. It is because $\eta$ has to vanish in the limits 
down to two and up to four dimensions separately. So significantly higher 
orders in $\epsilon$ as well as knowledge of the asymptotic behaviour of the 
series would be required even to get close to the three dimensional estimates
from the other methods.

A deeper underlying reason for this rests in the various theories which
populate a universality class. Viewing the class as driven by a fundamental
interaction at the Wilson-Fisher fixed point in $d$-dimensions, in the
neighbourhood of an even spacetime dimension there is a quantum field theory
which is renormalizable at that critical (even) dimension. Clearly it would
not be renormalizable at another critical dimension. However a different theory
would be relevant at that critical dimension but which will have the same
fundamental interaction. In the case of the Ising Gross-Neveu universality 
class the theory which is renormalizable in two dimensions is the Gross-Neveu
model of \cite{1}. By contrast in four dimensions the next theory in the tower
of theories at the Wilson-Fisher fixed point is what is termed the 
Gross-Neveu-Yukawa model, \cite{29}. The respective Lagrangians differ only in
the terms involving a scalar field. While we noted earlier that the Gross-Neveu
model is purely fermionic, the quartic self-interaction can be rewritten via a 
scalar auxiliary field $\sigma$ to produce a $\sigma \bar{\psi}^i \psi^i$
interaction where $\psi^i$ are the fermions. This interaction is common to the 
Gross-Neveu-Yukawa model but the $\sigma$ field now has a canonical kinetic 
term and a scalar quartic self-interaction, \cite{29}. It is this field 
$\sigma$ which has anomalous dimension $\eta_\phi$. Structurally the 
Gross-Neveu-Yukawa model has many similarities to the weak sector of the 
Standard Model itself. As noted in \cite{9}, for instance, this has led to the 
hope that certain phase transitions in graphene could mimic various symmetry 
breakings in the Standard Model itself such as chiral or spontaneous symmetry 
breaking. Currently experimental data from such graphene transitions are far 
from being available but the potential to have a simple laboratory to test 
Standard Model related phase transitions is a fascinating prospect. Other 
variations of the Gross-Neveu universality class are therefore becoming 
important to study so that precise theoretical predictions will be available.

Given the current sparcity of theoretical critical exponent estimates for the
chiral Heisenberg Gross-Neveu model it is crucial to bring the analysis up to
the same level of precision as that of the Ising Gross-Neveu universality
class. By this we mean that the amount of precise data of the two classes as 
represented in Figures $1$, $2$ and $3$ of \cite{9} should be the same for the 
different techniques used. This is the purpose of this article where we will 
compute the critical exponents $\eta$, $\eta_\phi$ and $1/\nu$ to the same
order in the large $N$ expansion as those in the Ising universality class. This
will be achieved by applying the critical point large $N$ formalism developed 
by Vasiliev et al in \cite{30,31,32} for the $O(N)$ nonlinear $\sigma$ model or
equivalently the $O(N)$ $\phi^4$ theory. Both these theories lie in the same 
universality class and the exponents of \cite{30,31,32} are expressed as 
functions of $d$. That original approach was later extended to the Ising 
Gross-Neveu universality class in a series of articles, 
\cite{22,23,24,25,26,27,28}, and it is on a subset of these papers that the
large $N$ computations presented here are based. For instance, for the most
part we will analyse the basic $2$-point functions of the chiral Heisenberg 
Gross-Neveu theory by solving the skeleton Schwinger-Dyson equations
algebraically in the limit as one approaches the $d$-dimensional 
Wilson-Fisher fixed point. While this will determine $\eta$, $\eta_\phi$ and
$1/\nu$ to $O(1/N^2)$, the original method given in \cite{32} pointed the way 
to determining $\eta$ at $O(1/N^3)$. This was extended to the Ising Gross-Neveu
class in \cite{24} and independently in \cite{28}. Therefore, we will apply 
what we now term the large $N$ conformal bootstrap method to the chiral
Heisenberg Gross-Neveu universality class. Given modern usage of the term 
conformal bootstrap method to mean a largely numerical technique based on 
solving the crossing symmetric $n$-point functions with manifest conformal 
symmetry, we have appended the term large $N$ to the naming of the earlier 
technique given in \cite{32}. The work of \cite{32} was inspired by the direct 
three dimensional conformal bootstrap construction of Parisi, \cite{33},
upon which the modern bootstrap, \cite{34,35,36,37}, is also based.

The advantage of the approach of \cite{30,31,32} is that the expression for
$\eta$ at $O(1/N^3)$ as well as the other exponents at $O(1/N^2)$ are 
determined as exact functions of $d$. It is important to appreciate that this 
arbitrary spacetime dimension is {\em not} the same as the one used for
dimensional regularization in perturbation theory. In \cite{30,31,32} the 
skeleton Schwinger-Dyson equations are analytically and not dimensionally 
regularized. Therefore the exponents derived in the large $N$ method correspond
to the exponents of the true theory at the Wilson-Fisher fixed point in 
arbitrary (non-integer) dimensions. Having said this the $\epsilon$ expansion 
of the large $N$ exponents around two or four dimensions are in exact agreement
with the $\epsilon$ expansion of the perturbative critical renormalization 
group functions, evaluated at the fixed point, of the underlying theory with a 
critical dimension of two or four respectively. One benefit of having the 
exponents as functions of $d$ is that we will be able to plot their behaviour 
in $2$~$<$~$d$~$<$~$4$ dimensions which will provide an independent insight 
into whether the dependence on $d$ is consistent with that given by the
functional renormalization group approach of \cite{9} which used a sharp 
regulator. While three dimensional estimates are clearly and ultimately of 
physical interest the non-integer dimension structure of a field theory in 
effect is indicating the properties of the underlying universality class. In 
the context of the Standard Model these simple Ising Gross-Neveu and chiral
Heisenberg Gross-Neveu classes may be pointing us to a novel way of viewing
four dimensional physics. For instance, the observation that supersymmetry can 
emerge at a fixed point in a multicoupling quantum field theory,
\cite{38,39,40}, may be directing us to a new era of beyond the Standard Model
analyses. In other words the couplings of the relevant operators with Standard 
Model symmetries may flow to a fixed point which is accessed by an effective 
quantum field theory in the short term. However this could be superseded by a 
new one in much the same way as the Lagrangian introducing the $W$ and $Z$ 
vector bosons was eventually constructed. Therefore, exploring this new chiral 
Heisenberg Gross-Neveu universality class and understanding it in the graphene 
context offers new exciting avenues to explore.

The article is organized as follows. The following section introduces the 
chiral Heisenberg Gross-Neveu universality class together with the basic large 
$N$ formalism which will be used to extract the critical exponents to several 
orders in $d$-dimensions. In section $3$ the exponent $\eta$ is determined at 
$O(1/N^2)$ and the evaluation of $\eta_\phi$ and $1/\nu$ to the same order are 
provided in sections $4$ and $5$ respectively. While these computations in 
essence follow the same critical point approach using the skeleton 
Schwinger-Dyson equations, the computation of $\eta$ at $O(1/N^3)$ uses the 
large $N$ conformal bootstrap method which is outlined in section $6$. The 
analysis of our results is given in section $7$ where estimates are shown for 
the case of graphene and plots of the large $N$ exponents in $d$-dimensions are
given for the Ising, XY and chiral Heisenberg Gross-Neveu models. Finally, we 
give our conclusions in section $8$.

\sect{Background.}

The chiral Heisenberg Gross-Neveu model (cHGN) is a generalization of the 
original two dimensional $O(N)$ or $SU(N)$ Gross-Neveu model of \cite{1} but 
where the $4$-point interaction is modified to include the three $SU(2)$ Pauli
spin matrices $\sigma^a$ where $a$~$=$~$1$, $2$ and $3$. The Lagrangian is 
\begin{equation}
L^{\mbox{\footnotesize{cHGN}}} ~=~ i \bar{\psi}^i \partialslash \psi^i ~+~
\frac{g^2}{2} \left( \bar{\psi}^i \sigma^a \psi^i \right)^2 
\label{lagchgnd2}
\end{equation}
where $1$~$\leq$~$i$~$\leq$~$N$ and $g$ is the dimensionless coupling constant.
Throughout our large $N$ computations we will use the spinor trace convention 
that $\mbox{Tr} I$~$=$~$2$ rather than $4$ which is ordinarily used when 
extending four dimensional perturbation theory to three dimensions. In other 
words to circumvent this convention we can rewrite our definition of $N$ using 
\begin{equation}
N ~=~ \frac{1}{2} d_\gamma N
\label{trcon}
\end{equation}
where $d_\gamma$ corresponds to the trace convention of the $\gamma$-matrices,
\cite{9}. The theory (\ref{lagchgnd2}) is only perturbatively renormalizable in
two dimensions. In four dimensions a quartic fermion interaction is 
non-renormalizable but (\ref{lagchgnd2}) can be reformulated in terms of an 
auxiliary field $\tilde{\pi}^a$ 
\begin{equation}
L^{\mbox{\footnotesize{cHGN}}} ~=~ i \bar{\psi}^i \partialslash \psi^i ~+~
g \tilde{\pi}^a \bar{\psi}^i \sigma^a \psi^i  ~-~ 
\frac{1}{2} \tilde{\pi}^a \tilde{\pi}^a ~. 
\label{lagchgn}
\end{equation}
In this formulation one can connect the original quartic interaction in the two
dimensional theory with one in four dimensions which we will term the chiral
Heisenberg Gross-Neveu-Yukawa (cHGNY) theory which has the Lagrangian
\begin{equation}
L^{\mbox{\footnotesize{cHGNY}}} ~=~ i \bar{\psi}^i \partialslash \psi^i ~+~
\frac{1}{2} \partial_\mu \tilde{\pi}^a \partial^\mu \tilde{\pi}^a ~+~
g_1 \tilde{\pi}^a \bar{\psi}^i \sigma^a \psi^i ~+~ 
\frac{1}{24} g_2^2 \left( \tilde{\pi}^a \tilde{\pi}^a \right)^2 ~. 
\label{lagchgny}
\end{equation}
The sharing of the common Yukawa interaction in (\ref{lagchgn}) and 
(\ref{lagchgny}) is related to the fact that these theories are in the same
universality class at their respective Wilson-Fisher fixed points. The bosonic 
sectors are different but they play a role in ensuring each Lagrangian is
renormalizable in their respective critical dimensions. While both theories lie
in the same universality class there is an underlying universal theory within
which one can carry out computations in the large $N$ expansion. Specifically
$d$-dimensional critical exponents can be determined to several orders in
$1/N$, when $N$ is large, using methods based on \cite{30,31,32}. These 
exponents encode all orders perturbative information on renormalization group 
functions in $d$-dimensions and hence contain information on all theories, such
as (\ref{lagchgn}) and (\ref{lagchgny}), in the universality class in their 
critical dimension. The Lagrangian for the application of the large $N$ method
to the underlying universal theory is 
\begin{equation}
L ~=~ i \bar{\psi}^i \partialslash \psi^i ~+~
\pi^a \bar{\psi}^i \sigma^a \psi^i  ~-~ \frac{1}{2g} \pi^a \pi^a
\label{laglargen}
\end{equation}
where the field $\tilde{\pi}^a$ has been rescaled so that there is no coupling
constant in the interaction. The method developed in \cite{30,31,32} relies on 
the fact that at criticality it is the interaction which drives the dynamics at
all scales. In comparison to (\ref{lagchgny}) there is no quartic interaction. 
This is not an issue since the structure of the universal theory as such 
generates the requisite information corresponding to the critical
renormalization group functions of (\ref{lagchgny}) via $\pi^a$ $4$-point
subgraphs in Feynman diagrams. This has been elucidated in \cite{40} in the 
context of the large $\Nf$ expansion of Quantum Chromodynamics (QCD) where 
$\Nf$ is the number of massless quark flavours. Equally for other theories in 
the tower of this universality class the higher $\pi^a$ $n$-point subgraphs 
will contain the vertex structure of the higher $n$-point interactions relevant
to the theory in a particular higher critical dimension. 

The large $N$ critical point formalism of \cite{30,31,32} exploits properties
of the renormalization group at criticality to provide a method of solving the
Schwinger-Dyson equations algebraically. In particular in the approach to a
fixed point there is no scale aside from the correlation length. So, for
instance, the propagators take an asymptotic scaling form whose behaviour is
governed solely by the full dimension of the field. Specifically in coordinate
space the scaling forms of the fermion and boson fields of (\ref{laglargen}) 
are, \cite{22}, 
\begin{equation}
\psi (x) ~ \sim ~ \frac{A\xslash}{(x^2)^\alpha} ~~~,~~~
\pi(x) ~ \sim ~ \frac{C}{(x^2)^\gamma}
\label{asform}
\end{equation}
at leading order in the limit as $x$~$\to$~$0$ where we use the name of the
field for each form. The quantities $A$ and $C$ are $x$-independent amplitudes
and $\alpha$ and $\gamma$ are the full dimensions of the actual fields in the
$d$-dimensional Lagrangian. These comprise the canonical dimension and the
anomalous dimension. The former is deduced from ensuring that the dimension of
each term in the Lagrangian is consistent with the action being dimensionless
and we define  
\begin{equation}
\alpha ~=~ \mu ~+~ \half \eta ~~~,~~~ \gamma ~=~ 1 ~-~ \eta ~-~ \chi_\pi
\end{equation}
in parallel with \cite{22} where $\eta$ is the anomalous critical exponent of 
the fermion. The anomalous dimension of $\pi^a$ includes $\eta$ and is derived 
from the Yukawa interaction where $\chi_\pi$ is the anomalous dimension of the 
Yukawa vertex itself. Here we use the convention of \cite{30,31} that the 
spacetime dimension $d$ is written as $d$~$=$~$2\mu$ for shorthand. In order to
be able to compare with the critical exponent of the analogous bosonic field of
\cite{12,14} we note that the corresponding exponent, $\eta_\phi$, is related 
to that of the $\pi^a$ field here by  
\begin{equation}
\eta_\phi ~=~ 4 ~-~ 2 \mu ~-~ 2 ( \eta ~+~ \chi_\pi ) ~.
\end{equation}
In terms of the connection with the renormalization group functions, $\eta$ 
corresponds to the fermion anomalous dimension evaluated at the Wilson-Fisher 
fixed point. As there is only one interaction the amplitudes $A$ and $C$ always
appear in the same combination which we will denote by $y$~$=$~$A^2C$. It like 
all the critical exponents will only depend on $\mu$ and $N$ at criticality. 
Hence they can be expanded in powers of $1/N$ via 
\begin{equation}
\eta(\mu) ~=~ \sum_{n=1}^\infty \frac{\eta_n(\mu)}{N^n} ~~~,~~~
y(\mu) ~=~ \sum_{n=1}^\infty \frac{y_n(\mu)}{N^n} 
\label{expexp}
\end{equation}
and a similar notation will be used for other exponents. In (\ref{expexp}) and 
similar expressions one can restore other spinor trace conventions by replacing
$N$ with the definition (\ref{trcon}). As we will be providing various critical
exponents we need to recall the background formalism for those. While 
(\ref{asform}) gives the dominant behaviour of the propagators in the approach 
to criticality corrections to scaling can be included and for the propagators 
these take the form, \cite{22},
\begin{equation}
\psi(x) ~\sim~ \frac{A\xslash}{(x^2)^\alpha} \left[ 1 + A^\prime(x^2)^\lambda
\right] ~~~,~~~
\pi(x) ~\sim~ \frac{C}{(x^2)^\gamma} \left[ 1 + C^\prime(x^2)^\lambda
\right] 
\label{ascorr}
\end{equation}
where an independent exponent $\lambda$ governs the scaling of the correction
with $A^\prime$ and $C^\prime$ the respective associated $x$-independent
amplitudes. In principle $\lambda$ can be any exponent but for the moment we 
will let it correspond to the critical slope of the $\beta$-function of the
theory whose critical dimension in this universality class is two and hence
corresponds to (\ref{lagchgnd2}) or (\ref{lagchgn}). The connection with the 
exponents of \cite{14} is that $1/\nu$~$=$~$2\lambda$ and we note that the 
leading order canonical dimension of $\lambda$ is 
$\lambda_0$~$=$~$\mu$~$-$~$1$. 

One of the main methods to determine the explicit values of the exponents to
several orders in $1/N$ is to systematically solve order by order the skeleton
Schwinger-Dyson $2$-point function for both fields in the approach to
criticality. This will also include the corrections to scaling. In order to do  
this one needs the asymptotic scaling forms of the respective $2$-point
functions. These are derived from the propagator forms by inverting the 
momentum space propagators. To do this we use the Fourier transform given in
\cite{30,31} which is  
\begin{equation}
\frac{1}{(x^2)^\alpha} ~=~ \frac{a(\alpha)}{2^{2\alpha}} \int_k
\frac{e^{ikx}}{(k^2)^{\mu-\alpha}}
\label{fourier}
\end{equation}
where
\begin{equation}
a(\alpha) ~=~ \frac{\Gamma(\mu-\alpha)}{\Gamma(\alpha)} ~.
\end{equation}
This produces the coordinate space $2$-point function asymptotic scaling forms
\begin{eqnarray}
\psi^{-1}(x) & \sim & \frac{r(\alpha-1)\xslash}{A(x^2)^{2\mu-\alpha+1}}
\left[ 1 - A^\prime s(\alpha-1)(x^2)^\lambda \right] \nonumber \\
\pi^{-1}(x) & \sim &\frac{p(\gamma)}{C(x^2)^{2\mu-\gamma}}
\left[ 1 - C^\prime q(\gamma) (x^2)^\lambda \right]
\label{astwo}
\end{eqnarray}
which include the scaling corrections. The amplitudes differ from the 
respective ones in the propagator forms and the various functions of the 
exponents are given by
\begin{eqnarray}
p(\gamma) &=& \frac{a(\gamma-\mu)}{a(\gamma)} ~~~,~~~
r(\alpha) ~=~ \frac{\alpha p(\alpha)}{(\mu-\alpha)} \nonumber \\
q(\gamma) &=& \frac{a(\gamma-\mu+\lambda)a(\gamma-\lambda)}
{a(\gamma-\mu)a(\gamma)} ~~~,~~~ 
s(\alpha) ~=~ \frac{\alpha(\alpha-\mu)q(\alpha)}{(\alpha-\mu+\lambda)
(\alpha-\lambda)} ~.
\end{eqnarray}
In summary we have introduced the basic structure of the propagators and
$2$-point functions for the large $N$ critical point analysis. This is 
sufficient to solve the skeleton Schwinger-Dyson equation at $O(1/N^2)$ and
determine $d$-dimensional expressions for $\eta_1$, $\eta_2$, $\chi_{\pi\,1}$,
$\lambda_1$ and $\lambda_2$. The values of $\chi_{\pi\,2}$ and $\eta_3$ are
deduced from solving $3$-point functions at criticality and the formalism for 
each will be discussed later. 

{\begin{figure}[hb]
\begin{center}
\includegraphics[width=11cm,height=3.5cm]{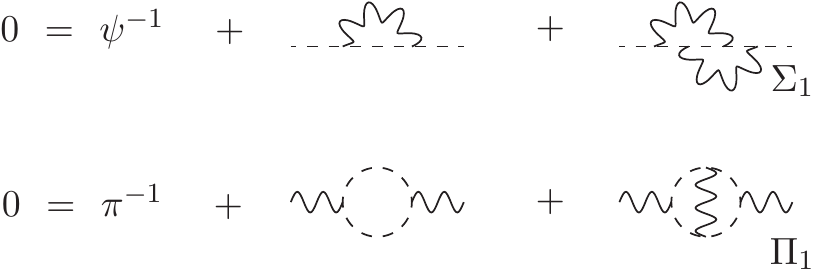}
\end{center}
\caption{$O(1/N^2)$ corrections to the skeleton Schwinger-Dyson $2$-point
functions used to determine $\eta_1$ and $\eta_2$.}
\end{figure}}

\sect{Evaluation of $\eta_2$.}

As aspects of the solution of the $2$- and $3$-point functions for each of the
various exponents are common we illustrate these features in more detail for
the derivation of $\eta_2$. The first stage for this is to represent the 
skeleton Schwinger-Dyson $2$-point functions of Figure $1$ as algebraic 
equations. In coordinate space to $O(1/N^2)$, which is sufficient to deduce
$\eta_1$ and $\eta_2$, we have 
\begin{eqnarray}
0 &=& r(\alpha-1) ~+~ 3 y Z_V^2 (x^2)^{\chi_\pi+\Delta}  ~-~ 
3 y^2 Z_V^4 \Sigma_1 (x^2)^{2\chi_\pi+2\Delta} ~+~ 
O \left( \frac{1}{N^3} \right) \nonumber \\
0 &=& p(\gamma) ~+~ 2 N y Z_V^2 (x^2)^{\chi_\pi+\Delta} ~+~ 
N y^2 Z_V^4 \Pi_1 (x^2)^{2\chi_\pi+2\Delta} ~+~ O \left( \frac{1}{N^2} \right)
\label{etasd}
\end{eqnarray}
where we have substituted for the asymptotic scaling forms of the inverse
propagators. The {\em values} of the respective two loop integrals $\Sigma_1$
and $\Pi_1$ are present and these not only depend on $N$ through the presence
of the exponents on the internal lines but also on the regularization $\Delta$.
This is introduced into the critical theory by shifting the anomalous dimension
of the vertex via, \cite{30,31}, 
\begin{equation}
\chi_\pi ~ \rightarrow ~ \chi_\pi ~+~ \Delta
\end{equation}
corresponding to an analytic regularization. Both graphs are divergent and can 
be represented by  
\begin{equation}
\Sigma_1 ~=~ \frac{K_1}{\Delta} ~+~ \Sigma_1^\prime ~~~,~~~ 
\Pi_1 ~=~ \frac{L_1}{\Delta} ~+~ \Pi_1^\prime 
\label{lographs}
\end{equation}
where the $O(\Delta)$ terms of the integrals would only become relevant for the
$\eta_3$ derivation using this approach. As discussed in \cite{30,31} and 
\cite{41} the core vertex is renormalized and this feature is present via the 
vertex renormalization constant $Z_V$ which also depends on $N$ and $\Delta$. 
It can be written as a double expansion via 
\begin{equation}
Z_V ~=~ 1 ~+~ \sum_{l=1}^\infty \sum_{n=1}^l \frac{m_{ln}}{\Delta^n} 
\end{equation}
where the counterterms $m_{ln}$ are then expanded in powers of $1/N$
\begin{equation}
m_{ln} ~=~ \sum_{i=1}^\infty \frac{m_{ln,i}}{N^i} ~.
\end{equation}

Next an overall common factor of the length of the position vector $x^2$ has 
cancelled in (\ref{etasd}). However, from the dimensionality of the Feynman 
graphs there is a remnant $x^2$ dependence deriving from the vertex anomalous 
dimension and the regularization. As it stands (\ref{etasd}) is a formal 
representation of Figure $1$ but is not applicable in the scaling region, given
by $x^2$~$\to$~$0$, where it ought to be independent of the length scale $x^2$.
Equally the renormalization constant has to be determined as well as 
$\chi_\pi$. It turns out that all these issues are resolved together. To 
achieve this we formally substitute (\ref{lographs}) and expand the factors 
involving $x^2$ to the appropriate orders in $1/N$ and $\Delta$ neglecting 
terms $O(1/N^3)$ and $O(\Delta)$ respectively. To ensure that the algebraic 
Schwinger-Dyson equation is finite when the regularization is removed leads to 
\begin{equation}
m_{11,1} ~=~ -~ \frac{1}{4} y_1 L_1
\end{equation}
and the absence of $\ln x^2$ terms requires
\begin{equation}
\chi_{\pi \, 1} ~=~ -~ \frac{1}{2} y_1 L_1 ~.
\end{equation}
We have expressed these conditions in terms of the simple pole of the graph 
$\Pi_1$ rather than $\Sigma_1$. However similar relations emerge from the 
$\psi^i$ Schwinger-Dyson equation and these are not inconsistent since from 
explicit computation we have 
\begin{equation}
K_1 ~=~ -~ \frac{1}{2} L_1
\label{polereln}
\end{equation}
which ensures there are no ambiguous expressions for the vertex counterterm and
$\chi_\pi$. The resulting Schwinger-Dyson equations 
\begin{eqnarray}
0 &=& r(\alpha-1) ~+~ 3 y ~-~ 3 y^2 \Sigma_1^\prime ~+~ 
O \left( \frac{1}{N^3} \right) \nonumber \\
0 &=& p(\gamma) ~+~2 N y ~+~ N y^2 \Pi_1^\prime ~+~ 
O \left( \frac{1}{N^2} \right)
\label{etasdfin}
\end{eqnarray}
are now finite and independent of $x^2$ to the order in $1/N$ necessary to
deduce $\eta_1$ and $\eta_2$. 

To proceed one first concentrates on the leading order terms of both equations
and eliminates $y_1$ to produce
\begin{equation}
p(\gamma) ~=~ \frac{2}{3} N r(\alpha-1) ~. 
\end{equation}
Setting the leading order value for $\gamma$ and 
$\alpha$~$=$~$\mu$~$+$~$\eta_1/N$ produces 
\begin{equation}
\eta_1 ~=~ -~ \frac{3 \Gamma(2\mu-1)}{\mu\Gamma(1-\mu)\Gamma(\mu-1)
\Gamma^2(\mu)} ~.
\end{equation}
With this value established as well as that for the amplitude combination $y_1$
then $\eta_2$ is deduced by including the values of the finite parts of the two
loop graphs and eliminating the unknown $y_2$. First, with 
\begin{equation}
L_1 ~=~ \frac{4}{(\mu-1)\Gamma^2(\mu)} 
\end{equation}
from \cite{22} we find 
\begin{equation}
\chi_{\pi\,1} ~=~ -~ \frac{\mu}{3(\mu-1)} \eta_1
\end{equation}
which is needed for the next term in the expansion of $p(\gamma)$. Then using
\begin{equation}
\Pi_1^\prime ~=~ -~ \frac{4}{(\mu-1)^2\Gamma^2(\mu)} 
\end{equation}
from \cite{22}, where $\Sigma_1^\prime$ satisfies the same ratio with 
$\Pi_1^\prime$ as (\ref{polereln}), we find
\begin{equation}
\eta_2 ~=~ \left[ \frac{(2\mu-3)}{3(\mu-1)} \Psi(\mu) ~+~ 
\frac{(4\mu^2-6\mu+1)}{2\mu(\mu-1)^2} \right] \eta_1^2
\end{equation} 
where we use the shorthand notation
\begin{equation}
\Psi(\mu) ~=~ \psi(2\mu-1) ~-~ \psi(1) ~+~ \psi(2-\mu) ~-~ \psi(\mu)
\end{equation}
and $\psi(z)$~$=$~$d \ln \Gamma(z)/dz$ is the Euler $\psi$ function.

\sect{Evaluation of $\chi_{\pi\,2}$.}

With the determination of the exponents of the last section we have obtained 
the critical behaviour of the $\psi^i$-field to $O(1/N^2)$ and that of $\pi^a$ 
to $O(1/N)$. Since the asymptotic scaling forms of the inverse scaling 
functions, when there is correction to scaling, involve $\gamma$ then in order 
to find $\lambda$ at $O(1/N^2)$ we need to find $\chi_{\pi\,2}$ first. One way 
is to determine the corrections to the skeleton $2$-point functions at next 
order in $1/N$. However this would involve a large amount of computation which 
would not be necessary since there is an alternative. The extraction of 
$\chi_{\pi\,1}$ from ensuring there are no terms involving $\ln x^2$ in the 
limit to criticality is reminiscent of a similar way of extracting the 
renormalization constant for a mass $m$ by renormalizing that part of a 
$2$-point function in conventional perturbation theory which corresponds to the
wave function renormalization. By this we mean that in a renormalizable theory 
the one loop mass counterterm can be deduced from the two loop $2$-point wave 
function computation by ensuring that there are no non-renormalizable terms 
involving $\ln m^2$. While this is not the conventional way to extract such a 
one loop counterterm from a two loop evaluation it can be regarded in one way 
as a useful shortcut but more importantly as being consistent with the 
underlying renormalizability. The same process has in effect been played out in
the finding $\chi_\pi$ at $O(1/N)$ through the computation of $\eta$ at
$O(1/N^2)$. While our discussion in the previous section followed the original 
algorithm outlined in \cite{30,31}, the critical point large $N$ formalism was 
subsequently put in a parallel context to conventional perturbation theory in 
\cite{41,42}. The core aspects of perturbation theory are an ordering of the 
diagrams constituting a Green's function and a regularization to facilitate 
their evaluation. With the ordering, such as the power of the coupling 
constant, the renormalization constants are introduced by multiplicative 
rescaling and determined with respect to a subtraction criterion. The situation
developed in \cite{41,42} is completely parallel. Graphs are ordered with 
respect to the counting given by the power of $1/N$ with the regularization 
introduced by shifting the vertex anomalous dimension, \cite{30,31}. One major 
difference is that the residues of the poles in the renormalization constants 
are functions of the spacetime dimension $d$ which does {\em not} play a 
regularizing role. In this process the critical exponents are directly 
extracted in a renormalization group invariant way. In introducing the vertex 
renormalization constant $Z_V$ previously we were in effect following the $1/N$
renormalization formalism developed in \cite{41,42} which demonstrated that the
ordering of graphs with a suitable regularization in effect was a complete 
parallel to conventional perturbation theory. Moreover it was shown in 
\cite{41,42} that the vertex anomalous dimensions, as well as operator 
renormalization, could be extracted from the critical point evaluation of 
$n$-point functions for $n$~$\geq$~$3$. Therefore to deduce $\chi_{\pi\,2}$ we 
follow that approach.

{\begin{figure}[ht]
\begin{center}
\includegraphics[width=2cm,height=2.0cm]{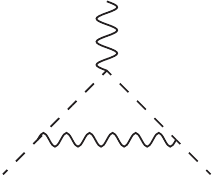}
\end{center}
\caption{$O(1/N)$ graph to determine $\chi_{\pi\,1}$.}
\end{figure}}

The first stage is to reproduce the value of $\chi_{\pi\,1}$ which requires the
evaluation of the graph of Figure $2$. In this instance we use the momentum 
space asymptotic scaling forms for the propagators which are 
\begin{equation}
\psi(p) ~\sim~ \frac{\tilde{A}\pslash}{(p^2)^{\mu-\alpha+1}} ~~~,~~~
\pi(p) ~\sim~ \frac{\tilde{C}}{(p^2)^{\mu-\gamma}}  
\label{ascorrmom}
\end{equation}
where $p$ is the momentum and $\tilde{A}$ and $\tilde{C}$ are the amplitudes
in momentum space. However the combination 
$\tilde{y}$~$=$~$\tilde{A}^2 \tilde{C}$ will always arise in the graphs. Its
$O(1/N^2)$ value is known via the Fourier transform (\ref{fourier}) and we
already have 
\begin{equation}
\tilde{y}_1 ~=~ -~ \frac{y_1}{(\mu-1)\Gamma(\mu)} ~~~,~~~ 
\tilde{y}_2 ~=~ -~ \frac{1}{(\mu-1)\Gamma(\mu)} \left[ y_2 ~-~
\frac{y_1 \eta_1}{(\mu-1)} \right] 
\end{equation}
where from earlier calculations we can deduce
\begin{equation}
y_1 ~=~ \frac{\mu}{6} \Gamma^2(\mu) \eta_1 ~~~,~~~ 
y_2 ~=~ \left[ \frac{\mu(2\mu-3)}{18(\mu-1)} \Psi(\mu) ~+~ 
\frac{\mu(5\mu-6)}{18(\mu-1)^2} \right] \Gamma^2(\mu) \eta_1^2 ~.
\end{equation}
The approach to find the corrections to $\chi_\pi$ is similar to that of the
$2$-point function. Once the graphs have been evaluated in momentum space to
the finite part in $\Delta$ the contribution to the respective term of the 
exponent in $1/N$ is given by the coefficient of the $\ln p^2$ part. Such terms
have to be absent in the limit to the critical point which then fixes the 
unknown exponent. Therefore following the prescription given in \cite{41} we 
find the same expression for $\chi_{\pi\,1}$ as before. 

{\begin{figure}[ht]
\begin{center}
\includegraphics[width=12cm,height=7.5cm]{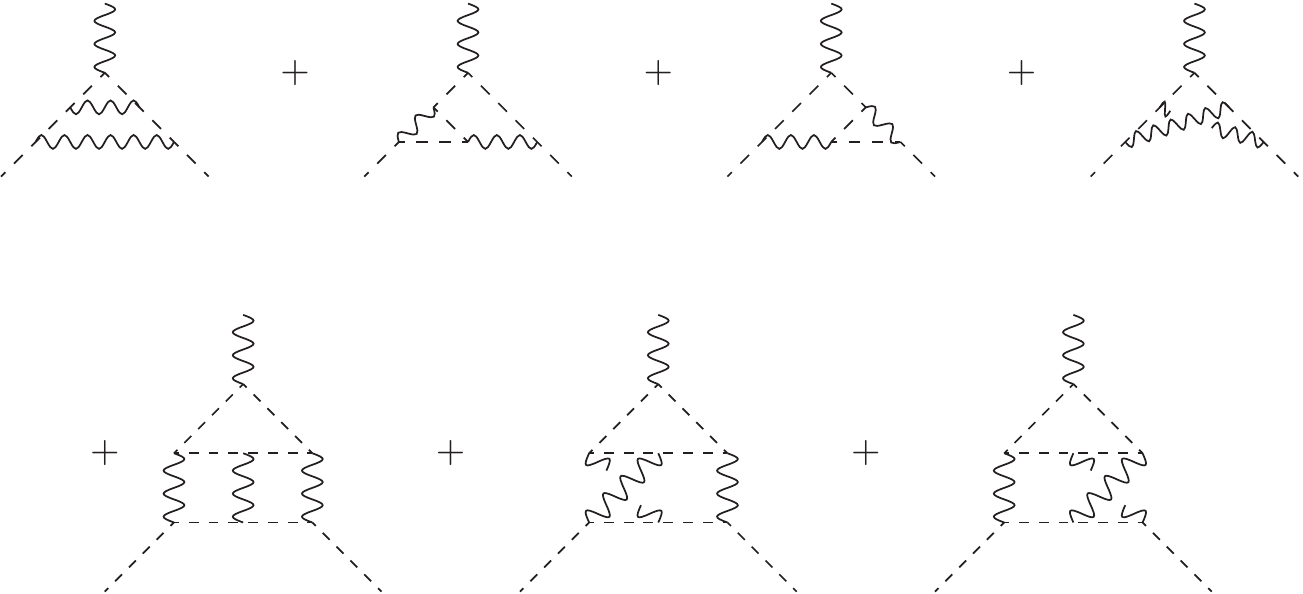}
\end{center}
\caption{$O(1/N^2)$ graphs to determine $\chi_{\pi\,2}$.}
\end{figure}}

The computation to deduce $\chi_{\pi\,2}$ is more involved. The main part is
the inclusion of the $O(1/N^2)$ graphs of Figure $3$. These were evaluated in
\cite{23} for the parallel calculation of the vertex critical exponent in the
$O(N)$ Gross-Neveu model. So for (\ref{laglargen}) the values of the master
integrals are appended with the corresponding group theory factor. In addition
there are next order contributions from the one loop graph of Figure $2$. This 
is because of the presence of $\eta$ and $\chi_\pi$ in the power of each of the 
propagators. In addition there are vertex counterterms from $Z_V$ for each 
vertex which have to be included in the same graph. They relate to the subgraph
divergences arising from the first three graphs of the top row in Figure $3$. 
Piecing all the relevant contributions together and isolating the $\ln p^2$ 
part we find 
\begin{equation}
\chi_{\pi\,2} ~=~ \left[ -~ \frac{\mu(2\mu-3)}{9(\mu-1)^2} \Psi(\mu) ~-~ 
\frac{\mu^2}{3\mu(\mu-1)} \Theta(\mu) ~-~ 
\frac{[ 14 \mu^3 - 37 \mu^2 + 31 \mu - 3]}{9(\mu-1)^2} \right] \eta_1^2 ~.
\end{equation}
where 
\begin{equation}
\Theta(\mu) ~=~ \psi^\prime(\mu) ~-~ \psi^\prime(1) ~.
\end{equation}
Equipped with $\chi_{\pi\,2}$ then both field anomalous dimensions are now 
available to the same order in $1/N$.

\sect{Evaluation of $\lambda_2$.}

The next stage in the evaluation of the large $N$ critical exponents again
mimics that of conventional perturbation theory in that with the wave function
anomalous dimensions at $O(1/N^2)$ one can establish the $\beta$-function to
the same order. As noted earlier this is achieved by considering corrections to
scaling and evaluating the corresponding $2$-point Schwinger-Dyson equations of
Figure $1$ to the next order. There are various aspects of extracting the
expansion for $\lambda$ which is not straightforward. Basic features are best
illustrated by considering the equation for $\psi^i$ which can be represented
by  
\begin{eqnarray}
0 &=& r(\alpha-1) \left[ 1 - A^\prime s(\alpha-1)(x^2)^\lambda \right] ~+~ 
3 y Z_V^2 (x^2)^{\chi_\pi+\Delta} 
\left[ 1 + \left( A^\prime + C^\prime \right) (x^2)^\lambda \right] 
\nonumber \\
&& -~ 3 y^2 Z_V^4 (x^2)^{2\chi_\pi+2\Delta} 
\left[ \Sigma_1 + \left( \Sigma_{1A} A^\prime + \Sigma_{1C} C^\prime 
\right) (x^2)^\lambda \right] ~+~ O \left( \frac{1}{N^3} \right) 
\end{eqnarray}
prior to renormalization. The two loop graph denoted by $\Sigma_1$ in Figure 
$1$ has been expanded to include the values where there is a correction to 
scaling in the $\psi^i$ and $\pi^a$ fields. These are $\Sigma_{1A}$ and
$\Sigma_{1C}$ respectively and their values have been computed in \cite{27}.
The key difference with the inclusion of the corrections is that terms do not
have the same dependence on $x^2$. This means that the equation decouples into
two pieces. One is the piece which contributes to the determination of $\eta$ 
while the other involves the unknown $\lambda$ and is 
\begin{eqnarray}
0 &=& -~ r(\alpha-1) s(\alpha-1) A^\prime ~+~ 
3 y Z_V^2 (x^2)^{\chi_\pi+\Delta} \left[ A^\prime + C^\prime \right]
\nonumber \\
&& -~ 3 y^2 Z_V^4 (x^2)^{2\chi_\pi+2\Delta} \left[ \Sigma_{1A} A^\prime 
+ \Sigma_{1C} C^\prime \right] ~+~ O \left( \frac{1}{N^3} \right) ~.
\label{psilam}
\end{eqnarray}
Unlike the equation which determines $\eta$ the associated correction 
amplitudes are present linearly. If one ignores the contribution from the two
loop graphs then the first two terms are the relevant ones for finding the
value of $\lambda_1$. However there is a new feature in comparison with the
corresponding equation to determine $\eta$. This is to do with the large $N$
dependence in that the first term is $O(1)$ while the second one is $O(1/N)$ 
which has an impact on the structure for the $\pi^a$ Schwinger-Dyson equation 
contributing to $\lambda$.  

Turning to the second equation of Figure $1$ we can represent it in a similar
way to that of the first equation and have 
\begin{eqnarray}
0 &=& p(\gamma) \left[ 1 - C^\prime q(\gamma)(x^2)^\lambda \right] ~+~ 
2 N y Z_V^2 (x^2)^{\chi_\pi+\Delta} \left[ 1 + 2 A^\prime (x^2)^\lambda \right] 
\nonumber \\
&& +~ N y^2 Z_V^4 (x^2)^{2\chi_\pi+2\Delta} 
\left[ \Pi_1 + \left( \Pi_{1A} A^\prime + \Pi_{1C} C^\prime \right) 
(x^2)^\lambda \right] ~+~ O \left( \frac{1}{N^2} \right) ~.
\end{eqnarray}
The values of the two loop graphs with corrections to scaling on the $\psi^i$ 
and $\pi^a$ lines are $\Pi_{1A}$ and $\Pi_{1C}$ respectively. Again the 
equation decouples into that which determines $\eta$ and 
\begin{eqnarray}
0 &=& -~ p(\gamma) q(\gamma) C^\prime ~+~ 
4 N y Z_V^2 (x^2)^{\chi_\pi+\Delta} A^\prime \nonumber \\
&& +~ N y^2 Z_V^4 (x^2)^{2\chi_\pi+2\Delta} \left[ \Pi_{1A} A^\prime
+ \Pi_{1C} C^\prime \right] ~+~ O \left( \frac{1}{N^2} \right) 
\label{pilam}
\end{eqnarray}
which is formally similar to (\ref{psilam}) and also includes the correction
amplitudes $A^\prime$ and $C^\prime$. Between these two equations at leading
order the only unknown is $\lambda_1$. The consistency equation which is formed
to solve for it is deduced by first representing (\ref{psilam}) and
(\ref{pilam}) as a matrix ${\cal M}$ given by
\begin{equation}
{\cal M} ~=~ \left(
\begin{array}{cc}
-~ r(\alpha-1) s(\alpha-1) & 3 y \\
4 N y & -~ p(\gamma) q(\gamma) ~+~ N y^2 \Pi^\prime_{1C} \\
\end{array}
\right)
\label{mat1}
\end{equation} 
after renormalization where ${}^\prime$ indicates the finite part of $\Pi_{1C}$
with respect to $\Delta$. It is worth noting that we have not retained all the 
terms in the Schwinger-Dyson equations in the matrix. This is because we want 
to concentrate on finding the value for $\lambda_1$ initially. Therefore we 
have retained the leading terms except for $\Pi^\prime_{1C}$ which would 
ordinarily be regarded as next order in the same way that the parent graph only
contributed to $\eta_2$ and not $\eta_1$. The reason why this graph cannot be 
omitted resides in the powers of $N$ in the matrix, \cite{27}, and the fact 
that the consistency equation is given by requiring $\det( {\cal M} )$~$=$~$0$. 
Therefore in computing the determinant the two resultant terms have to be the
same order in $1/N$. The complication arises from the scaling functions since
\begin{equation}
r(\alpha-1) s(\alpha-1) ~=~ O(1) ~~~,~~~ 
p(\gamma) q(\gamma) ~=~ O \left( \frac{1}{N} \right)  ~.
\end{equation} 
Therefore both terms of the $(22)$ element are the same order in $1/N$ and the
omission of $\Pi^\prime_{1C}$ would lead to an incorrect value of $\lambda_1$.
Therefore setting 
\begin{equation}
\Pi^\prime_{1C} ~=~ \Pi^\prime_{1C1} ~+~ \Pi^\prime_{1C2} \frac{1}{N} ~+~ 
O \left( \frac{1}{N^2} \right)
\label{pi1c}
\end{equation}
we have the leading order consistency equation 
\begin{equation}
12 y_1^2 ~-~ r(\alpha-1) s(\alpha-1) \left[ N p(\gamma) q(\gamma) ~-~
y_1^2 \Pi_{1C1} \right] ~=~ 0 
\end{equation}
where 
\begin{equation}
q(\gamma) ~=~ \frac{1}{2(\mu-1)N} \left[ \lambda_1 - \eta_1 - \chi_{\pi\,1}
\right] ~+~ O \left( \frac{1}{N^2} \right)
\end{equation}
to leading order. With, \cite{27},
\begin{equation}
\Pi^\prime_{1C1} ~=~ \frac{2}{(\mu-1)^2\Gamma^2(\mu)}
\end{equation}
we find 
\begin{equation}
\lambda_1 ~=~ -~ (2\mu-1) \eta_1 ~.
\end{equation}

{\begin{figure}[ht]
\begin{center}
\includegraphics[width=11cm,height=11cm]{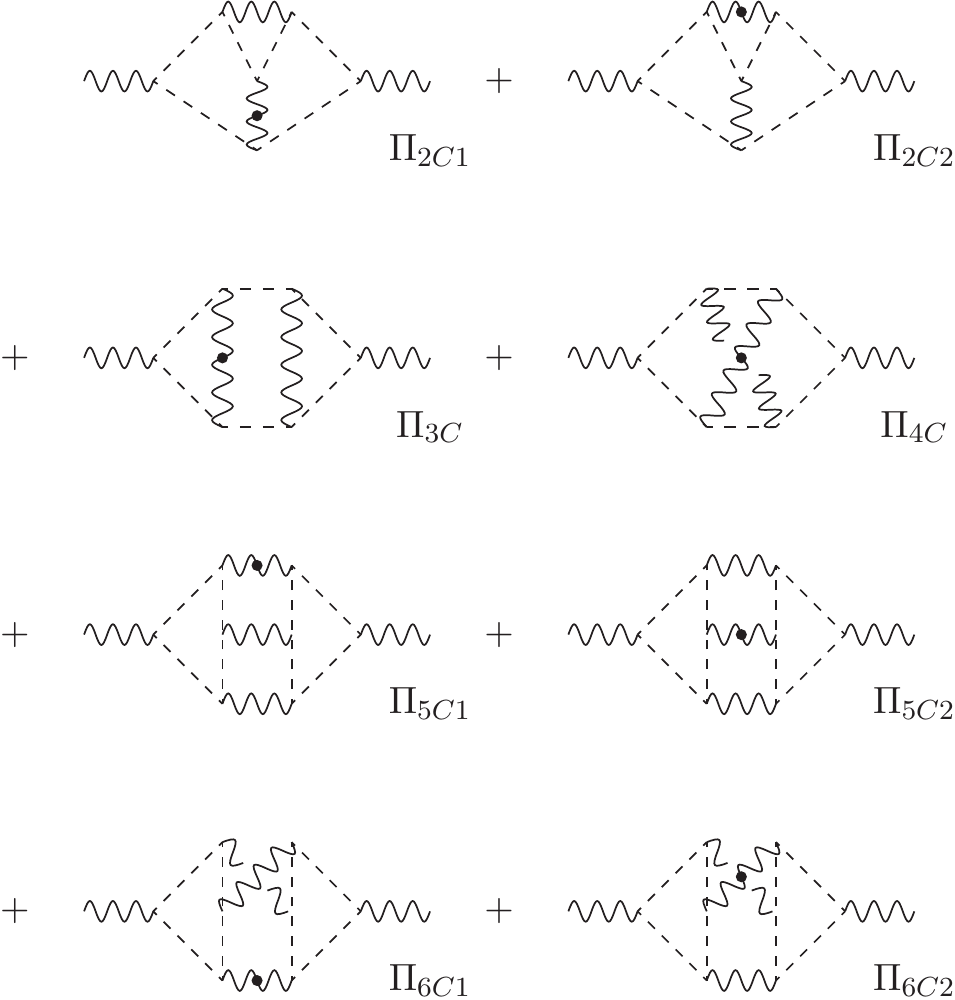}
\end{center}
\caption{Graphs for $O(1/N^2)$ correction to the $\pi^a$ skeleton 
Schwinger-Dyson $2$-point function to determine $\lambda_2$.}
\end{figure}}

Having outlined in detail the formalism and issues behind the derivation of the
leading order consistency equation for $\lambda_1$ we extend the analysis to
the next order in $1/N$. Virtually all aspects of the Schwinger-Dyson equations
discussed so far are sufficient to formally deduce the correction to the
consistency equation defined by $\det( {\cal M} )$~$=$~$0$. For instance, the
contributions from the two loop graph of the $\psi^i$ equation in Figure $1$
need to be included. However, for the $\pi^a$ equation there are two main
additions. The first is that the appearance of $\Pi_{1C}$ at leading order
means that the next term in its large $N$ expansion needs to be included as 
noted in (\ref{pi1c}) and the correction has the value, \cite{27}, 
\begin{equation}
\Pi^\prime_{1C2} ~=~ \frac{2}{(\mu-1)^2\Gamma^2(\mu)} \left[
\frac{3}{2} (\mu-1) \left[ \eta_1 - \chi_{\pi\,1} - \lambda_1 \right]
\left[ \Theta(\mu) + \frac{1}{(\mu-1)^2} \right] - \frac{2}{(\mu-1)} \eta_1
\right] ~.
\end{equation}
Also as another direct consequence of the presence of $\Pi_{1C}$ at leading the
$O(1/N^2)$ corrections to the $\pi^a$ Schwinger-Dyson equation have to be
included. These are illustrated in Figure $4$ where the dot on a $\pi^a$ line
denotes which line has the correction to scaling term. The values of the
integrals were given in \cite{27} but in this case the group factors have to be
appended. We use a labelling of the graphs which is parallel to that used in
\cite{27}. Moreover their contribution has to be included in the extension of
(\ref{pilam}). If we denote the sum of the contributions from the graphs of 
Figure $4$ by $\Pi_{2C}$
then the extension of (\ref{mat1}) is
\begin{equation}
{\cal M} ~=~ \left(
\begin{array}{cc}
-~ r(\alpha-1) s(\alpha-1) ~+~ 3 y & 3 y ~-~ 3 y^2 \Sigma^\prime_{1C} \\
4 N y ~+~ N y^2 \Pi^\prime_{1A} & 
-~ p(\gamma) q(\gamma) ~+~ N y^2 \Pi^\prime_{1C} ~+~ \Pi_{C2} \\
\end{array}
\right) ~. 
\end{equation} 
Summing all the explicit contributions to $\Pi_{C2}$ we have 
\begin{eqnarray}
\Pi_{C2} &=& \left[ 
\frac{2\mu[ 8 \mu^2 - 16 \mu + 7]}{3(\mu-1)^2(\mu-2)^2 \eta_1} ~-~ 
\frac{ \mu^2 [ 44 \mu^3 - 192 \mu^2 + 287 \mu - 151 ]}{18(\mu-1)^4(\mu-2)^2}
\right. \nonumber \\
&& \left. ~-~ \frac{ \mu^2 [ 2\mu - 3 ]}{3(\mu-1)^2(\mu-2)} \left[ \Phi(\mu)
+ \Psi^2(\mu) \right] ~+~ \frac{ \mu^2 [ 8 \mu - 11 ]}{3(\mu-1)^2(\mu-2)}
\Theta(\mu) \right. \nonumber \\
&& \left. ~+~ \frac{\mu^2[16\mu^3-82\mu^2+151\mu-95]}{9(\mu-1)^3(\mu-2)^2} 
\right] y_1 \eta_1^2 ~.
\end{eqnarray} 
With this value together with the corrections to the asymptotic scaling 
functions it is straightforward to solve $\det( {\cal M} )$~$=$~$0$ at 
$O(1/N^2)$ and find
\begin{eqnarray}
\lambda_2 &=& \left[ \frac{\mu^2[6\mu^2-3\mu-8]}{6(\mu-1)(\mu-2)} 
\Theta(\mu) ~-~ \frac{2\mu^2(2\mu-3)}{3(\mu-1)(\mu-2)} \left[
\Phi(\mu) + \Psi^2(\mu) \right] ~+~
\frac{2\mu[8\mu^2-16\mu+7]}{3(\mu-1)(\mu-2)^2\eta_1} \right. \nonumber \\
&& \left. ~+~ 
\frac{[72 \mu^8 - 604 \mu^7 + 1960 \mu^6 - 3060 \mu^5 + 2151 \mu^4 - 146 \mu^3 
- 621 \mu^2 + 288 \mu - 36]}{18 \mu (\mu - 1)^3 (\mu - 2)^2}
\right. \nonumber \\
&& \left. ~-~ \frac{(2\mu-3)[18\mu^5-95\mu^4+161\mu^3-86\mu^2-20\mu+12]}
{9(\mu-1)^2(\mu-2)^2} \Psi(\mu) \right] \eta_1^2
\end{eqnarray}
where 
\begin{equation}
\Phi(\mu) ~=~ \psi^\prime(2\mu-1) ~-~ \psi^\prime(2-\mu) ~-~ 
\psi^\prime(\mu) ~+~ \psi^\prime(1) ~.
\end{equation}
We note that the coefficients in the Taylor expansion of this function near two
dimensions, together with $\Psi(\mu)$ and $\Theta(\mu)$, all involve $\zeta_n$ 
for integer $n$~$\geq$~$3$ where $\zeta_z$ is the Riemann zeta function. 

Finally we close this section by briefly mentioning that we have repeated the
procedure at leading order to find the critical exponent which relates to one
part of the $\beta$-functions of (\ref{lagchgny}). To do this instead of using 
the correction to scaling of (\ref{ascorr}) we use the alternative correction  
\begin{equation}
\psi(x) ~\sim~ \frac{A\xslash}{(x^2)^\alpha} \left[ 1 + A^\prime(x^2)^\omega
\right] ~~~,~~~
\pi(x) ~\sim~ \frac{C}{(x^2)^\gamma} \left[ 1 + C^\prime(x^2)^\omega
\right] 
\end{equation}
where $\omega_0$~$=$~$\mu$~$-$~$2$. The consistency equation for $\omega_1$ is
formally the same as that for $\lambda_1$ except that $\lambda$ is replaced by 
$\omega$. The same reordering at leading order occurs and the analogous value
to $\Pi^\prime_{1C1}$ is required. Denoting this by $\Pi^\prime_{1C1\,\omega}$ 
we note that, \cite{43}, 
\begin{equation}
\Pi^\prime_{1C1\,\omega} ~=~ -~ 
\frac{2(\mu^2-4\mu+2)}{(\mu-1)^2(\mu-2)\Gamma^2(\mu)} 
\end{equation}
and find
\begin{equation}
\omega_1 ~=~ -~ \frac{2(2\mu-1)(2\mu-3)(\mu-2)}{3(\mu-1)} \eta_1 ~.
\end{equation}
As in \cite{43} this corresponds to one of the eigen-critical exponents of the 
$2$~$\times$~$2$ matrix of derivatives of the two $\beta$-functions of
(\ref{lagchgny}). While the $O(1/N^2)$ corrections are known to the 
(\ref{lagchgny}) $\beta$-functions, \cite{44}, the method used in that approach
differed from that used here. Specifically the relevant critical exponents were
determined by examining $3$-point functions. So the values of the master 
integrals computed in \cite{44} cannot be immediately translated to the 
extension of our $2$-point Schwinger-Dyson equation to next order in $1/N$.  

\sect{Large $N$ conformal bootstrap.}

The provision of $\lambda_2$ completes the evaluation of the three basic
exponents $\eta$, $\eta_\phi$ and $1/\nu$ to $O(1/N^2)$ in the large $N$ 
expansion. The next stage is to proceed to $O(1/N^3)$ which is possible in the
case of $\eta$. However this is not by the evaluation of the two $2$-point
function Schwinger-Dyson equations. While in principle one should be able to
extend the $\eta_2$ calculation it transpires that the evaluation of the higher
order graphs is not straightforward. Instead we follow the method developed in 
\cite{32} based on the earlier work of \cite{33} which we will term the large 
$N$ conformal bootstrap method. In this approach one analyses the skeleton 
Schwinger-Dyson equation for the $3$-point vertex using the dressed propagators
(\ref{asform}) but with additionally no vertex subgraphs. The focus therefore 
is on what would be called the primitive graphs in the $3$-point function. 
These are illustrated in Figure $5$ where there is a difference in the 
representation of the vertices in comparison with the earlier Figures. The dot 
at each vertex represents what is termed a conformal triangle, \cite{32,33}. In
other words the original vertex is replaced by a one loop triangle graph where 
the exponents of the new internal edges are at this stage arbitrary. Their 
values are fixed by the criterion that each new vertex in the triangle is
unique. A vertex is said to be unique if the sum of the exponents of the lines 
joining a $3$-point vertex is equal to the spacetime dimension $d$. The concept
of uniqueness was introduced in three dimensions in \cite{45} and extended to 
$d$-dimensions in \cite{30,31}. One consequence of uniqueness is that in
applying a conformal transformation to any of the graphs of Figure $5$
immediately reduces it to a $2$-point function. While this means that as these 
graphs stand they can be computed, the question still remains as to how to 
extract $\eta_3$. 

{\begin{figure}[ht]
\begin{center}
\includegraphics[width=11cm,height=7.5cm]{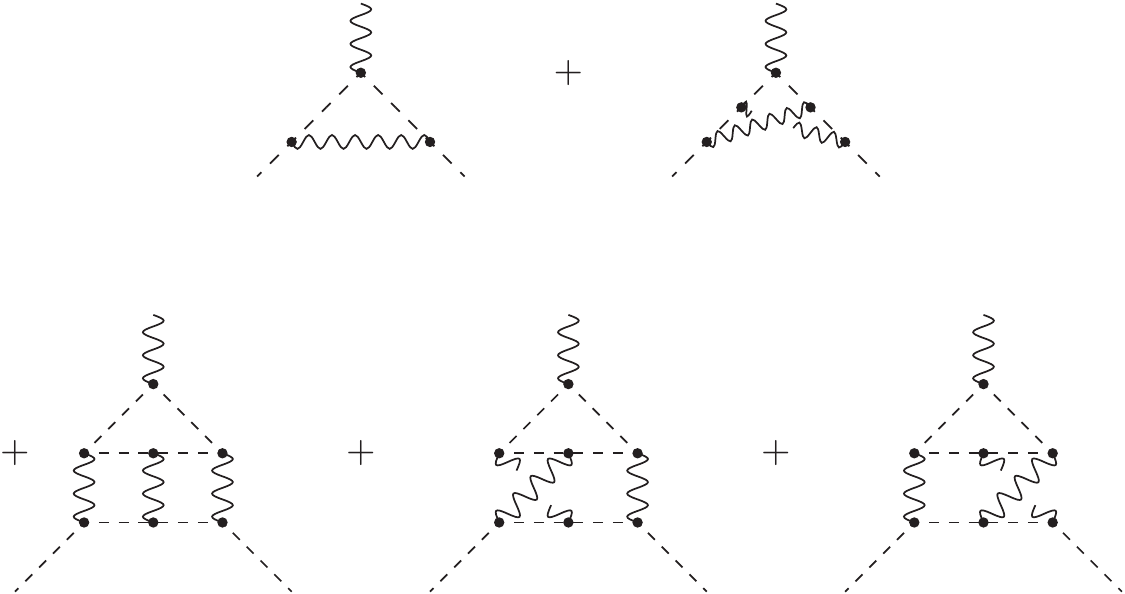}
\end{center}
\caption{Graphs contributing to the vertex function which determines $\eta_3$ 
using the large $N$ conformal bootstrap method.}
\end{figure}}

The original procedure to carry this out for the nonlinear $\sigma$ model in
$d$-dimensions was given in \cite{32} and involved extending the work of
\cite{33} which was specific to three dimensions. The first stage is to 
construct the consistency equations whose solution gives $\eta_3$. This has
been given in \cite{24,25,46} for the $O(N)$ Gross-Neveu model and that 
construction straightforwardly translates to the case of (\ref{laglargen}). The
only major difference aside from the different labelling of the exponents is to
append the factors deriving from the Pauli matrices in the vertex. Therefore we
focus on outlining the formalism. The key ingredient is the vertex function 
which is denoted by $V(y,\alpha,\gamma;\delta,\delta^\prime)$. Here $y$ is the 
same combination of amplitudes as before and the function depends on two 
regularizing parameters $\delta$ and $\delta^\prime$. These are required since 
in the derivation of the equation for $\eta$ given in \cite{24,25,32,46} there 
are divergent $2$-point functions. The divergences arise in the same context as 
those in the earlier $2$-point function analyses which required the 
introduction of the analytic regularization controlled by $\Delta$. As the 
conformal bootstrap also is a perturbative expansion in the vertex anomalous 
dimension the vertices of the graphs with conformal triangles also have to 
regularized. This is achieved by requiring that the external $\pi^a$ and one of
the external $\psi^i$ legs of the graphs in Figure $5$ have their dimension 
shifted by $\delta$ and $\delta^\prime$ respectively, \cite{24,25,46}.  

{\begin{figure}[hb]
\begin{center}
\includegraphics[width=11cm,height=7.5cm]{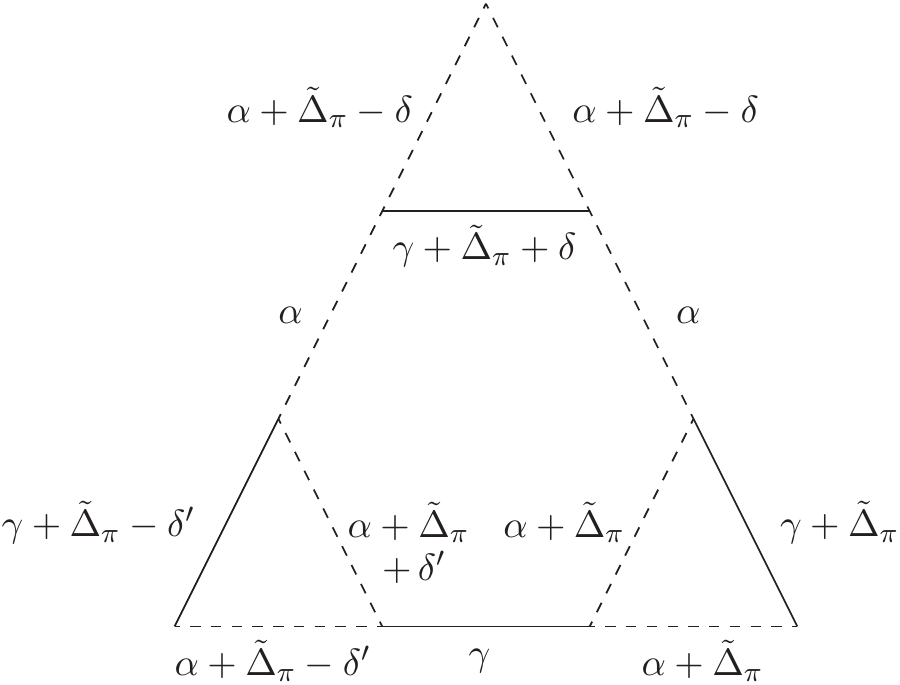}
\end{center}
\caption{Regularized leading order graph for conformal bootstrap construction.}
\end{figure}}

Having outlined features of the vertex function the formalism presented in
\cite{24,25,46} leads to the consistency equations. First the vertex function
$V(y,\alpha,\gamma;\delta,\delta^\prime)$ is defined by the sum of  
$\delta$-regularized graphs given in Figure $5$. The basic equation which in
effect determines the hidden amplitide of the vertex order by order in large
$N$ is 
\begin{equation}
1 ~=~ V(y,\alpha,\gamma;0,0) ~.
\label{cnfb1}
\end{equation}
In other words the sum of all the graphs is unity and it turns out that each
graph is finite when evaluated after applying conformal techniques. Therefore
the regularization for this was unnecessary, \cite{24,25,46}. However, the
graphs which contribute to the value of $\eta_3$ and equally to the lower order
exponents are divergent which means the regularization cannot be neglected in
their determination. Consequently the second consistency equation of the set is
\begin{eqnarray}
\frac{2N r(\alpha-1)}{3p(\gamma)} &=& \left.
\frac{ \left[ 1 + 2 \chi_\pi \frac{\partial ~}{\partial \delta^\prime}
V(y,\alpha,\gamma;\delta,\delta^\prime) \right]}
{ \left[ 1 + 2 \chi_\pi \frac{\partial ~}{\partial \delta}
V(y,\alpha,\gamma;\delta,\delta^\prime) \right]}
\right|_{\delta=\delta^\prime=0} 
\label{cnfb2}
\end{eqnarray}
where the same scaling functions as before are present. At leading order the
right hand side is unity which means that the same value for $\eta_1$ is
recovered as expected. At next order the one loop graph of Figure $5$ has to be
included. However, it has to be evaluated with the regularized external
vertices and conformal triangles. In order to illustrate the earlier points
about the graphs in the large $N$ conformal bootstrap construction we have
given the explicit allocation of exponents on the lines in Figure $6$ for the 
full graph corresponding to the one loop graph of Figure $5$. There for space
considerations we have set 
\begin{equation}
\chi_\pi ~=~ 2 \tilde{\Delta}_\pi ~.
\end{equation}
It is clear that there is a non-zero sum of the exponents at the top and bottom
left vertices which are proportional to $\delta$ and $\delta^\prime$ 
respectively. Equally the sum of all the exponents at internal vertices are
unique. The value of the graph in Figure $6$ has been determined in the context
of the usual Gross-Neveu model as a function of the exponents of that model,
\cite{46}. So that formal expression can be adapted to this computation as the 
group factor is separate for the graph in Figure $6$. Exporting that function 
and inserting it into (\ref{cnfb1}) and (\ref{cnfb2}) we recover the earlier 
value of $\eta_2$ which provides a useful check on our conformal bootstrap 
construction.

The final stage is to include the remaining graphs of Figure $5$ in the two
consistency equations. Again the values of the contributing integrals have 
been computed in \cite{28} in the context of the usual Gross-Neveu model. So 
those master integral values need only be decorated with the group factors for 
the specific case we are interested in. Unlike the one loop graph of Figure $5$
some of the higher order $\delta$-regularized graphs cannot be evaluated
completely. However for these cases it transpires that the difference
\begin{equation}
\left. \left[ \frac{\partial ~}{\partial \delta^\prime}
V(y,\alpha,\gamma;\delta,\delta^\prime) ~-~ \frac{\partial ~}{\partial \delta}
V(y,\alpha,\gamma;\delta,\delta^\prime) \right]
\right|_{\delta=\delta^\prime=0}
\end{equation}
can be determined. This is all that is necessary since this difference for the
higher order graphs is sufficient to deduce $\eta_3$ in the $1/N$ expansion of 
the right hand side of (\ref{cnfb2}). If we denote the contributions to the 
right hand side of (\ref{cnfb2}) from all the graphs in Figure $5$ except the 
one loop triangle by $V_2$ then
\begin{eqnarray}
V_2 &=& \left[ \frac{\mu^2(14\mu^2-37\mu+28)}{18(\mu-1)^3} \Psi(\mu) ~-~ 
\frac{\mu^2 (2\mu-5) (7\mu^2-18\mu+16)}{18(\mu-1)^4} \right. \nonumber \\
&& \left. ~-~ \frac{\mu^2 (\mu-16)}{36(\mu-1)^2} \Theta(\mu) ~-~
\frac{\mu^2}{6(\mu-1)} \Xi(\mu) \left[ \Theta(\mu) + \frac{1}{(\mu-1)^2} 
\right] \right] \eta_1^2 ~.
\end{eqnarray}
This involves a new function $\Xi(\mu)$ which is related to a particular two 
loop graph introduced in \cite{32} and denoted there by $I(\mu)$. In terms of 
$I(\mu)$ we have 
\begin{equation}
I(\mu) ~=~ -~ \frac{2}{3(\mu-1)} ~+~ \Xi(\mu)
\end{equation}
so that the expansion of $\Xi(\mu)$ near two dimensions only involves multiple 
zeta values, \cite{47}. For instance, the first few terms are
\begin{equation}
\Xi(1-\epsilon) ~=~ \frac{2}{3} \zeta_3 \epsilon^2 ~+~ \zeta_4 \epsilon^3 ~+~ 
\frac{13}{3} \zeta_5 \epsilon^4 ~+~ O(\epsilon^5) ~.
\end{equation}
In three dimensions the integral $I(\mu)$ is known exactly, \cite{32}, since
\begin{equation}
I(\threehalves) ~=~ 2 \ln 2 ~+~ \frac{3\psi^{\prime\prime}(\half)}{2\pi^2}
\end{equation}
which will be needed for our three dimensional estimates later. Finally 
including $V_2$ in (\ref{cnfb2}) and expanding the one loop contribution from
Figure $6$ to $O(1/N^2)$ we find that 
\begin{eqnarray}
\eta_3 &=& \left[ \frac{(2\mu-3)}{18(\mu-1)^2} 
\left[ \Phi(\mu) + 3 \Psi^2(\mu) \right] ~-~ 
\frac{[\mu^3+18\mu^2-21\mu+9]}{36(\mu-1)^2} \Theta(\mu) ~-~ 
\frac{\mu^2}{3(\mu-1)} \Theta(\mu) \Psi(\mu) \right. \nonumber \\
&& \left. ~-~ \frac{\mu^2}{6(\mu-1)^3} \Xi(\mu) ~-~ 
\frac{[14 \mu^7 - 15 \mu^6 - 26 \mu^5 - 77 \mu^4 + 324 \mu^3 - 297 \mu^2 
+ 90 \mu - 9]}{18\mu^2(\mu-1)^4} 
\right. \nonumber \\
&& \left. ~-~ \frac{[14\mu^5-37\mu^4-50\mu^3+228\mu^2-183\mu+27]}
{18\mu(\mu-1)^3} \Psi(\mu) ~-~ \frac{\mu^2}{6(\mu-1)} \Xi(\mu) \Theta(\mu) 
\right] \eta_1^3 
\end{eqnarray}
which completes the evaluation of all the critical exponents. To assist with
analyses we have provided an attached data file where electronic versions of
all the exponents computed here are given.  

\sect{Results.}

We devote this section to discussing our results and give estimates of critical
exponents in three dimensions. As a first stage we must indicate that all the
expressions derived here are consistent with known perturbation theory near
four dimensions. Recently the three and four loop renormalization group
functions have been provided for the chiral Heisenberg Gross-Neveu-Yukawa
theory, \cite{12,14}, which built on the early loop work of \cite{10,11}. 
Setting $d$~$=$~$4$~$-$~$2\epsilon$ in the $d$-dimensional exponents we find
\begin{eqnarray}
\left. \frac{}{} \eta \right|_{d=4-2\epsilon} &=& \left[ 3 \epsilon 
- \frac{9}{2} \epsilon^2 - \frac{9}{4} \epsilon^3
+ \frac{3}{8} [ 16 \zeta_3 - 3 ] \epsilon^4 + \frac{9}{16} \left[ 16 \zeta_4 
- 16 \zeta_3 - 1 \right] \epsilon^5 \right] \frac{1}{N} 
\nonumber \\
&& +~ \left[ -~ 3 \epsilon + \frac{99}{4} \epsilon^2 - \frac{303}{8} \epsilon^3
- \frac{3}{2} [ 16 \zeta_3 + 13 ] \epsilon^4 + 3 \left[ 49 \zeta_3 - 12 \zeta_4
+ 3 \right] \epsilon^5 \right] \frac{1}{N^2} 
\nonumber \\
&& +~ \left[ 3 \epsilon - \frac{153}{4} \epsilon^2 + \left[ 72 \zeta_3
- \frac{51}{4} \right] \epsilon^3
+ \frac{3}{16} [ 576 \zeta_4 + 3337 - 2104 \zeta_3 ] \epsilon^4 
\right. \nonumber \\
&& \left. ~~~~
+~ \frac{3}{32} \left[ 1536 \zeta_5 - 10645 - 6312 \zeta_4 + 2104 \zeta_3 
\right] \epsilon^5 \right] \frac{1}{N^3} ~+~ 
O \left( \epsilon^6; \frac{1}{N^4} \right) \nonumber \\
\left. \frac{}{} \eta_\phi \right|_{d=4-2\epsilon} &=& 2 \epsilon ~+~ 
\left[ - 2 \epsilon + 5 \epsilon^2 
+ \frac{1}{2} \epsilon^3 + \left[ - \frac{7}{4} - 4 \zeta_3 \right] \epsilon^4 
+ \frac{1}{8} \left[ 80 \zeta_3 - 48 \zeta_4 - 23 \right] \epsilon^5 \right] 
\frac{1}{N} \nonumber \\
&& +~ \left[ 2 \epsilon + \frac{69}{2} \epsilon^2 
+ \left[ 48 \zeta_3 - \frac{625}{4} \right] \epsilon^3 
+ \left[ 155 - 128 \zeta_3 + 72 \zeta_4 \right] \epsilon^4 
\right. \nonumber \\
&& \left. ~~~~
+~ 2 \left[ 73 \zeta_3 - 96 \zeta_4 + 48 \zeta_5 + 18 \right] \epsilon^5 
\right] \frac{1}{N^2} ~+~ O \left( \epsilon^6; \frac{1}{N^3} \right) 
\nonumber \\
\left. \frac{1}{\nu} \right|_{d=4-2\epsilon} &=& 2 ~-~ 2 \epsilon \nonumber \\
&& +~ \left[ - 18 \epsilon + 39 \epsilon^2 - \frac{9}{2} \epsilon^3 
+ \left[ - \frac{9}{4} - 36 \zeta_3 \right] \epsilon^4 ~+~ 
\frac{3}{8} \left[ 208 \zeta_3 - 144 \zeta_4 - 3 \right] \epsilon^5 \right] 
\frac{1}{N} \nonumber \\
&& +~ \left[ 438 \epsilon - \frac{3005}{2} \epsilon^2 
+ \left[ \frac{3721}{4} - 552 \zeta_3 \right] \epsilon^3 
+ \left[ 744 + 3608 \zeta_3 - 828 \zeta_4 + 1360 \zeta_5 \right] \epsilon^4 
\right. \nonumber \\
&& \left. ~~~~
+~ \frac{1}{2} \left[ 1760 \zeta_3^2 - 16532 \zeta_3 + 10824 \zeta_4 
- 17728 \zeta_5 + 6800 \zeta_6 - 365 \right] \epsilon^5 \right] 
\frac{1}{N^2} \nonumber \\
&& +~ O \left( \epsilon^6; \frac{1}{N^3} \right) ~.
\end{eqnarray}
Comparing with the results of \cite{10,12,13,14} and allowing for the different 
$\epsilon$ expansion convention we find exact agreement with the exponents of
\cite{10,12,13,14}. For future five loop computations we have also given the 
$O(\epsilon^5)$ terms. We have also checked that the $\epsilon$ expansion of 
$\omega_1$ near four dimensions is in agreement with one of the eigen-critical 
exponents of the Hessian defined by the derivatives of the two 
$\beta$-functions of \cite{14}. This check is similar to that carried out in
the Ising Gross-Neveu model \cite{43}. For similar reasons we give the 
$\epsilon$ expansion of the exponents in $d$~$=$~$2$~$-$~$2\epsilon$ to the 
same order. We have
\begin{eqnarray}
\left. \frac{}{} \eta \right|_{d=2-2\epsilon} &=& \left[ 3 \epsilon^2 
+ 3 \epsilon^3 + 3 \epsilon^4 + \left[ 3 + 6 \zeta_3 \right] \epsilon^5 \right]
\frac{1}{N} \nonumber \\
&& +~ \left[ -~ \frac{9}{2} \epsilon^2 - \frac{45}{2} \epsilon^3
- 36 \epsilon^4 + \left[ - 30 \zeta_3 - 45 \right] \epsilon^5 \right] 
\frac{1}{N^2} \nonumber \\
&& +~ \left[ -~ 9 \epsilon^2 + 24 \epsilon^3 + 159 \epsilon^4 
+ \left[ 396 - \frac{111}{2} \zeta_3 \right] \epsilon^5 \right] 
\frac{1}{N^3} ~+~ O \left( \epsilon^6; \frac{1}{N^4} \right) \nonumber \\
\left. \frac{}{} \eta_\phi \right|_{d=2-2\epsilon} &=& 2 ~+~ 2 \epsilon ~+~ 
\left[ - 2 \epsilon - 6 \epsilon^2 - 6 \epsilon^3 
+ \left[ - 6 - 4 \zeta_3 \right] \epsilon^4 
+ \left[ -~ 12 \zeta_3 - 6 \zeta_4 - 6 \right] \epsilon^5 \right] 
\frac{1}{N} \nonumber \\
&& +~ \left[ -~ 10 \epsilon - 3 \epsilon^2 + 23 \epsilon^3 
+ \left[ 78 - 44 \zeta_3 \right] \epsilon^4 
+ \left[ 36 \zeta_3 - 66 \zeta_4 + 96 \right] \epsilon^5 
\right] \frac{1}{N^2} \nonumber \\
&& +~ O \left( \epsilon^6; \frac{1}{N^3} \right) \nonumber \\
\left. \frac{}{} \frac{1}{\nu} \right|_{d=2-2\epsilon} &=& -~ 2 \epsilon ~+~ 
\left[ -~ 6 \epsilon^2 + 6 \epsilon^3 + 6 \epsilon^4 
+ \left[  6 - 12 \zeta_3 \right] \epsilon^5 \right] \frac{1}{N} \nonumber \\
&& +~ \left[ 8 \epsilon^2 + 9 \epsilon^3 
+ \left[ 67 + 90 \zeta_3 \right] \epsilon^4 
+ \left[ 240 \zeta_3 + 135 \zeta_4 - 178 \right] \epsilon^5 \right] 
\frac{1}{N^2} \nonumber \\
&& +~ O \left( \epsilon^6; \frac{1}{N^3} \right) ~.
\end{eqnarray}
Having established the consistency of our $d$-dimensional critical exponents
with known four dimensional perturbation theory it is a simple exercise to
determine the values in three dimensions. We have
\begin{eqnarray}
\left. \frac{}{} \eta \right|_{d=3} &=& \frac{4}{\pi^2N} ~+~ 
\frac{64}{3\pi^4N^2} ~+~ \frac{8 [ 378 \zeta_3 - 36 \pi^2 \ln(2)
- 45 \pi^2 - 332 ]}{9\pi^6N^3} ~+~ O \left( \frac{1}{N^4} \right)
\nonumber \\
\left. \frac{}{} \eta_\phi \right|_{d=3} &=& 1 ~+~ 
\frac{16[3\pi^2+16]}{3\pi^4N^2} ~+~ O \left( \frac{1}{N^3} \right)
\nonumber \\
\left. \frac{1}{\nu} \right|_{d=3} &=& 1 ~-~ \frac{16}{\pi^2N} ~+~
\frac{[ 144 \pi^2 + 1664 ]}{3\pi^4N^2} ~+~ O \left( \frac{1}{N^3} \right)
\end{eqnarray}
where the $O(1/N)$ correction to $\eta_\phi$ is zero or
\begin{eqnarray}
\left. \frac{}{} \eta \right|_{d=3} &=& \frac{0.4052845}{N} ~+~ 
\frac{0.219008}{N^2} ~-~ \frac{0.525197}{N^3} ~+~ 
O \left( \frac{1}{N^4} \right) \nonumber \\
\left. \frac{}{} \eta_\phi \right|_{d=3} &=& 1 ~+~ 
\frac{2.497169}{N^2} ~+~ O \left( \frac{1}{N^3} \right)
\nonumber \\
\left. \frac{1}{\nu} \right|_{d=3} &=& 1 ~-~ \frac{1.621139}{N} ~+~
\frac{10.557615}{N^2} ~+~ O \left( \frac{1}{N^3} \right)
\end{eqnarray}
numerically. 

{\begin{table}[ht]
\begin{center}
\begin{tabular}{|c||l|l|l|l|}
\hline
  & $1/\nu$ & $\eta_\phi$ & $\eta$ & $\nu$ \\
\hline
$\epsilon$ expansion $[2,2]$ Pad\'{e} \cite{14} & $0.6426$ & $0.9985$ & 
$0.1833$ & $-$ \\
$\epsilon$ expansion $[3,1]$ Pad\'{e} \cite{14} & $0.6447$ & $0.9563$ & 
$0.1560$ & $1.2352$ \\
Functional RG \cite{48} & $0.795$ & $1.032$ & $0.071$ & $1.26$ \\
Monte Carlo \cite{49} & $(0.98)$ & & $0.20(2)$ & $1.02(1)$ \\
Monte Carlo \cite{50} & $(1.19)$ & $0.70(15)$ & & $0.84(4)$ \\
Large $N$ & $0.8458$ & $1.1849$ & $0.1051$ & $1.1823$ \\
\hline
\end{tabular}
\end{center}
\begin{center}
{Table $1$. Comparison of large $N$ Pad\'{e} estimates with other methods for
$N$~$=$~$4$ for the chiral Heisenberg Gross-Neveu model.}
\end{center}
\end{table}}

Equipped with these three dimensional expressions we can now provide estimates 
for the exponents in the case which is of interest to graphene problems. 
Therefore recalling our convention for the spinor trace (\ref{trcon}) the value
of $N$ we use in order to compare with the numerical estimates of critical 
exponents of other methods for (\ref{laglargen}) is $N$~$=$~$4$. Given this we 
have determined numerical estimates for the three critical exponents using 
Pad\'{e} approximants. These are given in the last line of Table $1$ together 
with results quoted in \cite{14} by other approaches for comparison. In the 
table bracketed values for $1/\nu$ represent the value derived from $\nu$ which
was computed directly. For the large $N$ estimates we have quoted the $[1,2]$, 
$[0,2]$ and $[1,1]$ Pad\'{e} approximants for $\eta$, $\eta_\phi$ and $1/\nu$ 
respectively. In general the values for large $N$ are similar but larger than 
those of the functional renormalization group values of \cite{48} for 
$\eta_\phi$ and $1/\nu$. For $\eta$ the situation is much different with 
virtually no overlap for any of the different analyses. As \cite{14} also 
considered the XY Gross-Neveu model we have also provided the parallel results 
for $N$~$=$~$4$ in Table $2$ in order to compare results for the same methods 
in a different context. Again it is the case that the estimates for $\eta_\phi$
and $1/\nu$ are slightly larger than those of the functional renormalization 
group values in \cite{14} with again no clear consensus for $\eta$. Although it
is worth noting that the large $N$ estimate for $\eta$ is a $[1,1]$ Pad\'{e} 
since the value of $\eta_3$ for that model has not been computed.  

{\begin{table}[hb]
\begin{center}
\begin{tabular}{|c||l|l|l|}
\hline
  & $1/\nu$ & $\eta_\phi$ & $\eta$ \\
\hline
$\epsilon$ expansion $[2,2]$ Pad\'{e} \cite{14} & $0.840$ & $0.810$ & 
$0.117$ \\
$\epsilon$ expansion $[3,1]$ Pad\'{e} \cite{14} & $0.841$ & $0.788$ & 
$0.108$ \\
Functional RG \cite{51} & $0.862$ & $0.88$ & $0.062$ \\
Monte Carlo \cite{52} & $1.06(5)$ & $0.71(3)$ & \\
Large $N$ & $0.9026$ & $0.9023$ & $0.0872$ \\
\hline
\end{tabular}
\end{center}
\begin{center}
{Table $2$. Comparison of large $N$ Pad\'{e} estimates with other methods for
$N$~$=$~$4$ for the chiral XY Gross-Neveu model.}
\end{center}
\end{table}}

Perhaps a more instructive way of viewing the situation with the exponent 
estimates is to use a graphical representation motivated by the functional
renormalization group approach \cite{9}. One aspect of that method is that it 
is not limited to a discrete spacetime dimension. In other words the critical 
exponents can be determined as functions of $d$ and plots of the three
exponents in $2$~$<$~$d$~$<$~$4$ were given in \cite{9}. As our expansion
parameter, $1/N$, is dimensionless in $d$-dimensions it is possible to provide 
similar plots for the same exponents. By this we mean that we can plot the
various Pad\'{e} approximants used to obtain the estimates in Table $1$ for
the chiral Heisenberg Gross-Neveu model as functions of $d$. These are given
in Figure $7$ for $N$~$=$~$4$ except that the $[1,1]$ Pad\'{e} is used for 
$\eta$. This is because we do not have a closed analytic form for the 
$\Xi(\mu)$ as a function of $\mu$. Instead we have used the first two terms of
$\eta$. However in the plot of $\eta$ we have included the $[2,1]$ Pad\'{e}
estimate, indicated by an open circle, and the $[1,2]$ estimate indicated by a
solid circle. These are meant to guide roughly where the $d$-dimensional line 
would intersect if $\eta_3$ was known as an analytic function in 
$d$-dimensions. One interesting aspect of the three curves in Figure $7$ is 
that they are in good qualitative agreement with those of the right hand panels
of Figures $1$, $2$ and $3$ of \cite{9}. By this we mean that the shape in 
terms of concavity and convexity for $\eta_\phi$ and $1/\nu$ are very similar 
as well as the offset of the peak for $\eta$ which is not in the neighbourhood 
of $d$~$=$~$3$. Finally, in order to assist this comparison we have included 
similar plots for the XY Gross-Neveu model when $N$~$=$~$4$ and the ordinary or
Ising Gross-Neveu model for $N$~$=$~$8$ in Figures $8$ and $9$ respectively. 
The latter value of $N$ in that case is the parallel one to compare with 
\cite{9}. In Figure $8$ only the first two terms of $\eta$ were available which
why the Pad\'{e} estimate lies on the line. While the shapes of the plots for 
the Ising Gross-Neveu model are also qualitatively similar to those of \cite{9}
it is worth noting that the one for $\eta_\phi$ differs in concavity to that 
for the chiral Heisenberg Gross-Neveu model which is consistent with the 
functional renormalization group approach.  


{\begin{figure}[hb]
\begin{center}
\includegraphics[width=7cm,height=7cm]{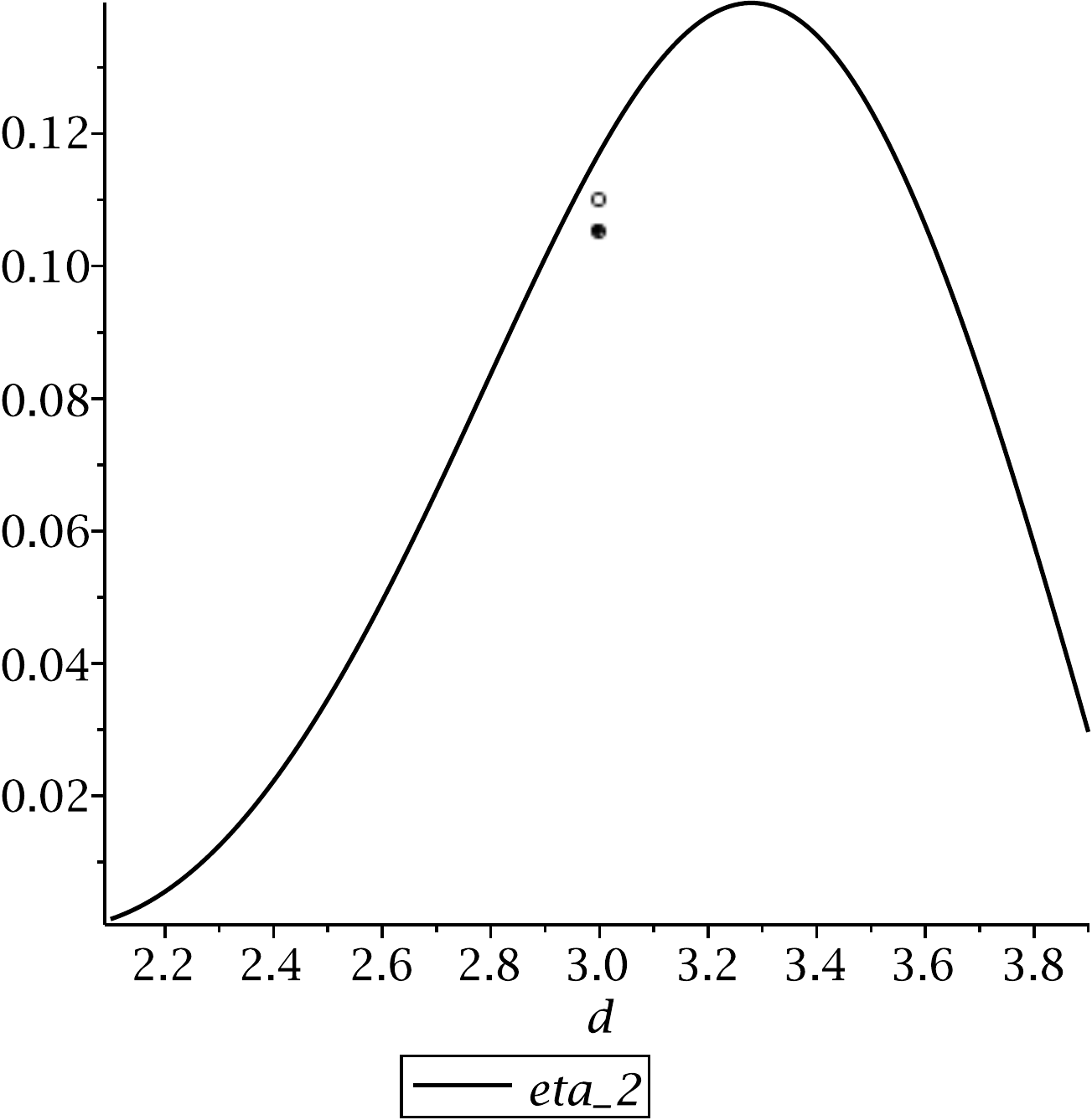}
\quad
\includegraphics[width=7cm,height=7cm]{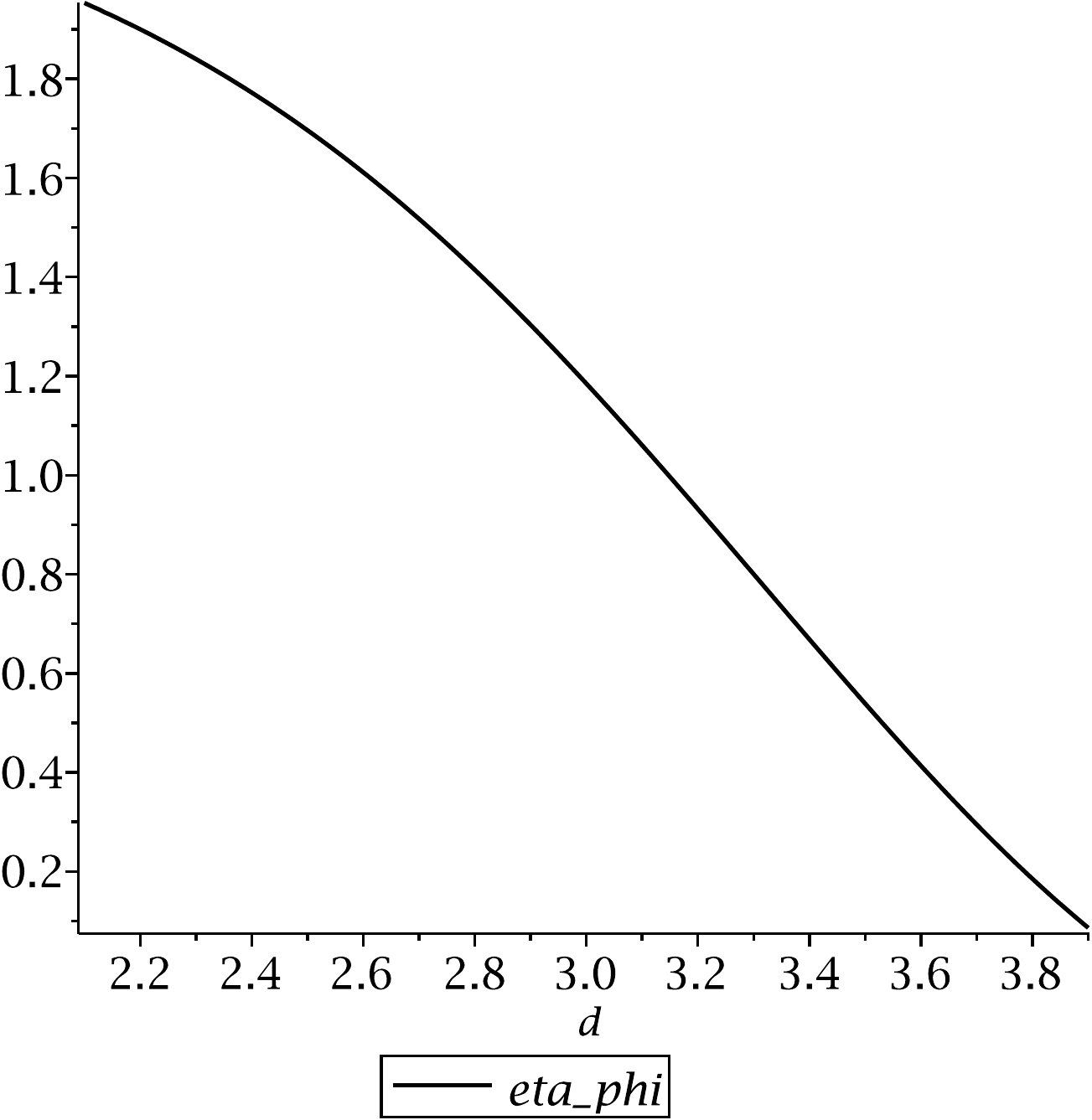} \\ 
\vspace{1.5cm}
\includegraphics[width=7cm,height=7cm]{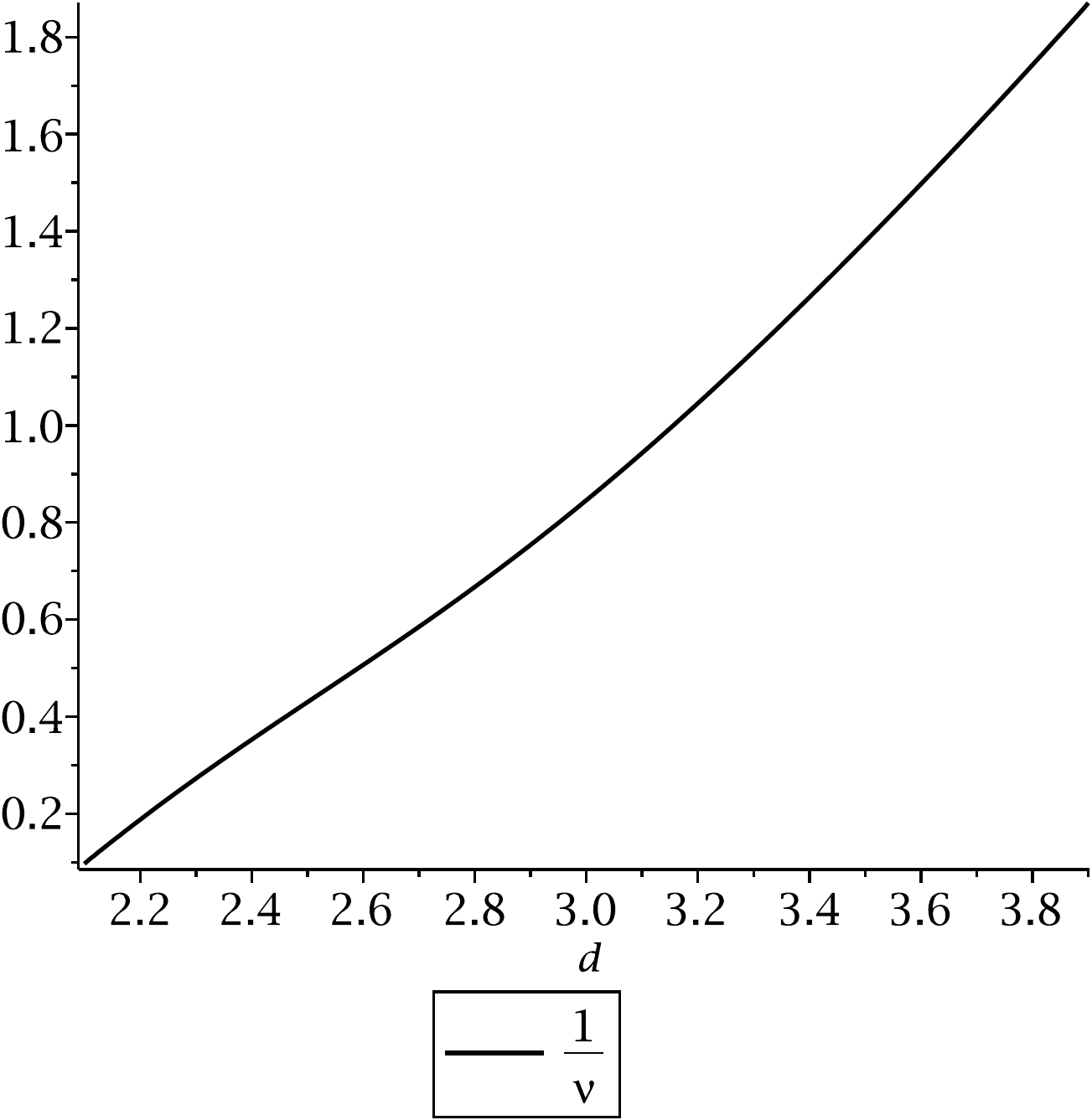}
\quad
\end{center}
\caption{Dependence of Pad\'{e} approximants to critical exponents $\eta$,
$\eta_\phi$ and $1/\nu$ for the chiral Heisenberg Gross-Neveu model when 
$N$~$=$~$4$. In the $\eta$ plot the open circle is the $[2,1]$ Pad\'{e} 
approximant using $\eta$ at $O(1/N^3)$ and the solid circle is the $[1,2]$ 
Pad\'{e} approximant.}
\end{figure}}

\clearpage 

{\begin{figure}[hb]
\begin{center}
\includegraphics[width=7cm,height=7cm]{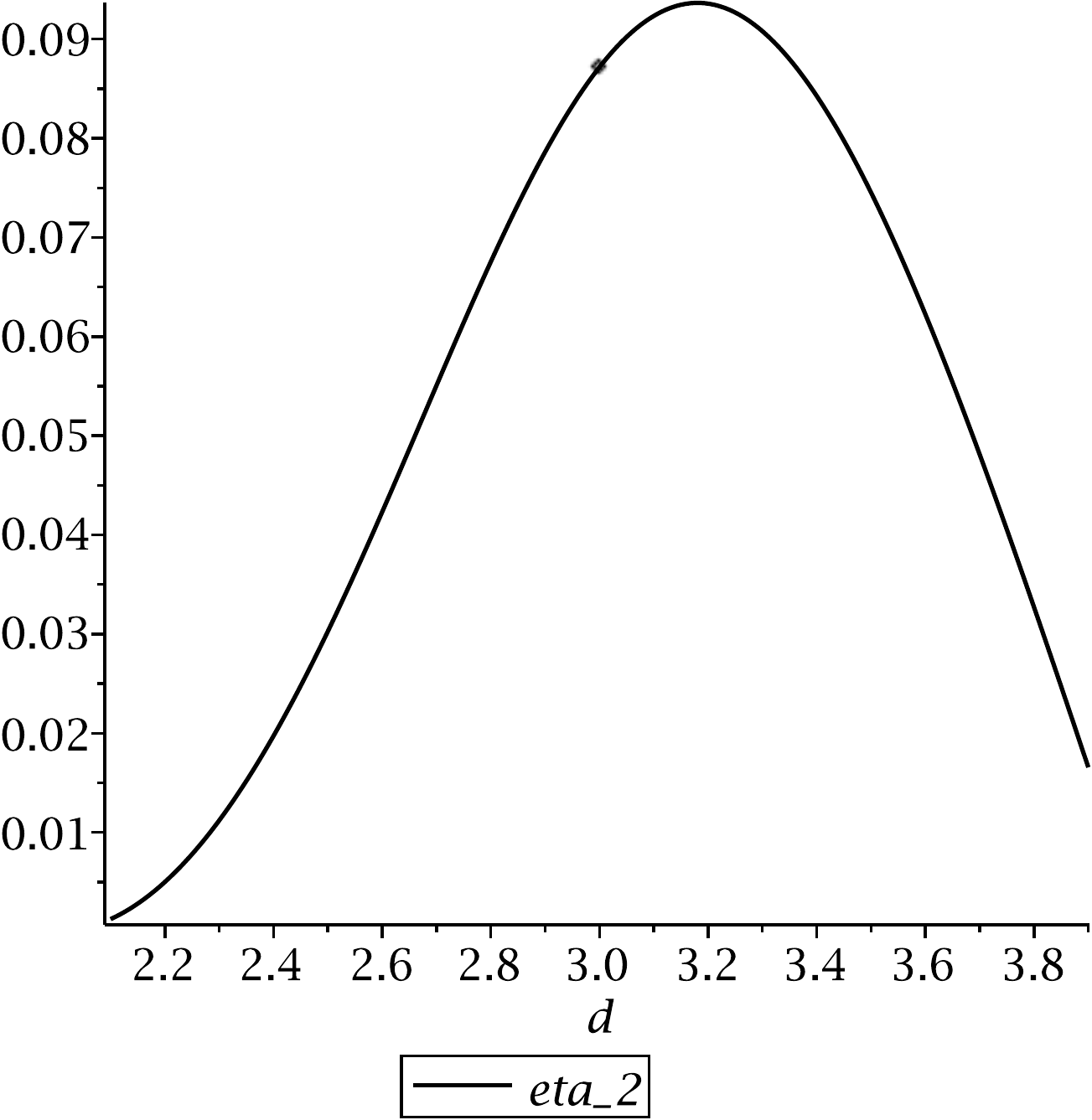}
\quad
\includegraphics[width=7cm,height=7cm]{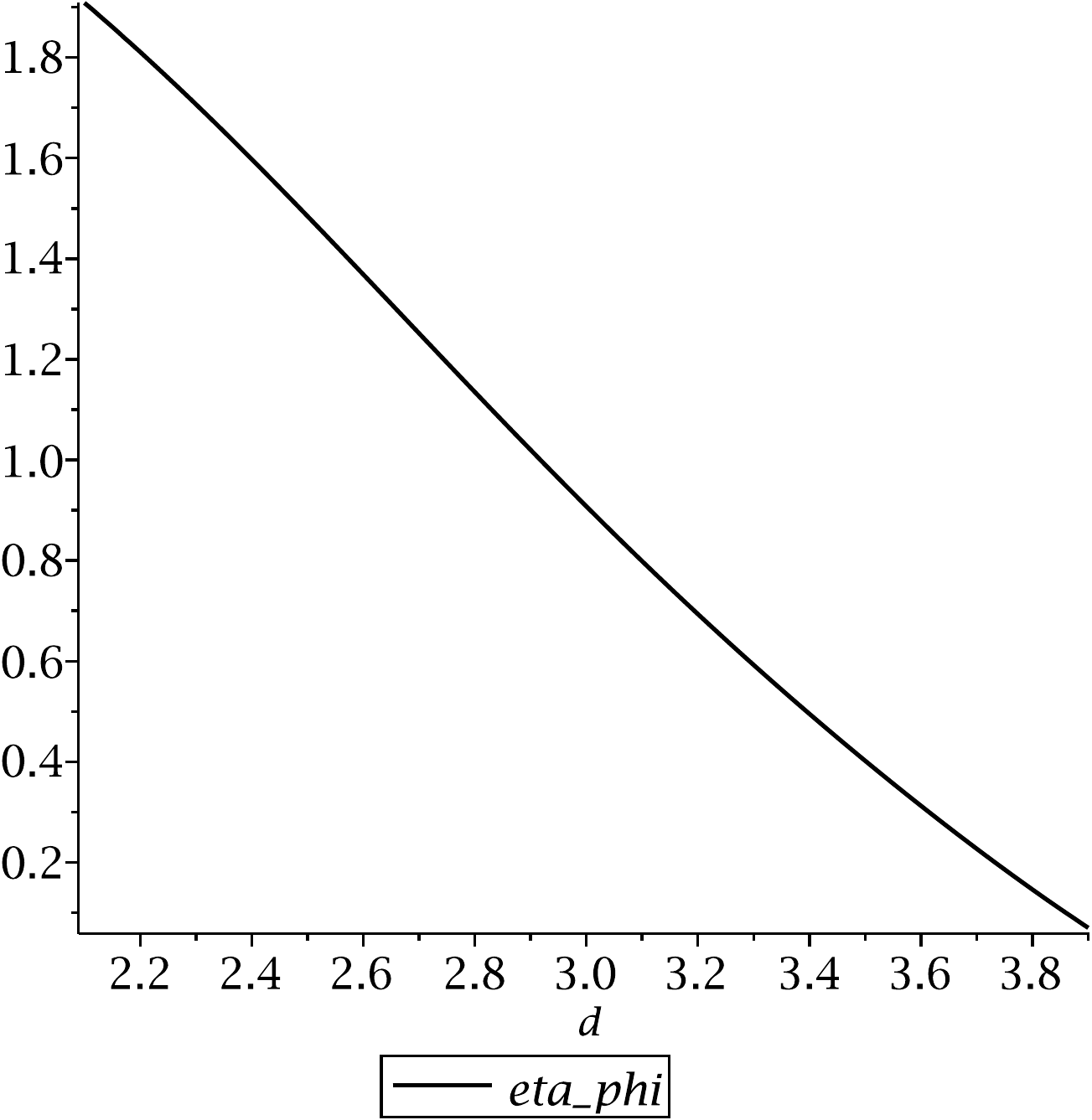} \\ 
\vspace{1.5cm}
\includegraphics[width=7cm,height=7cm]{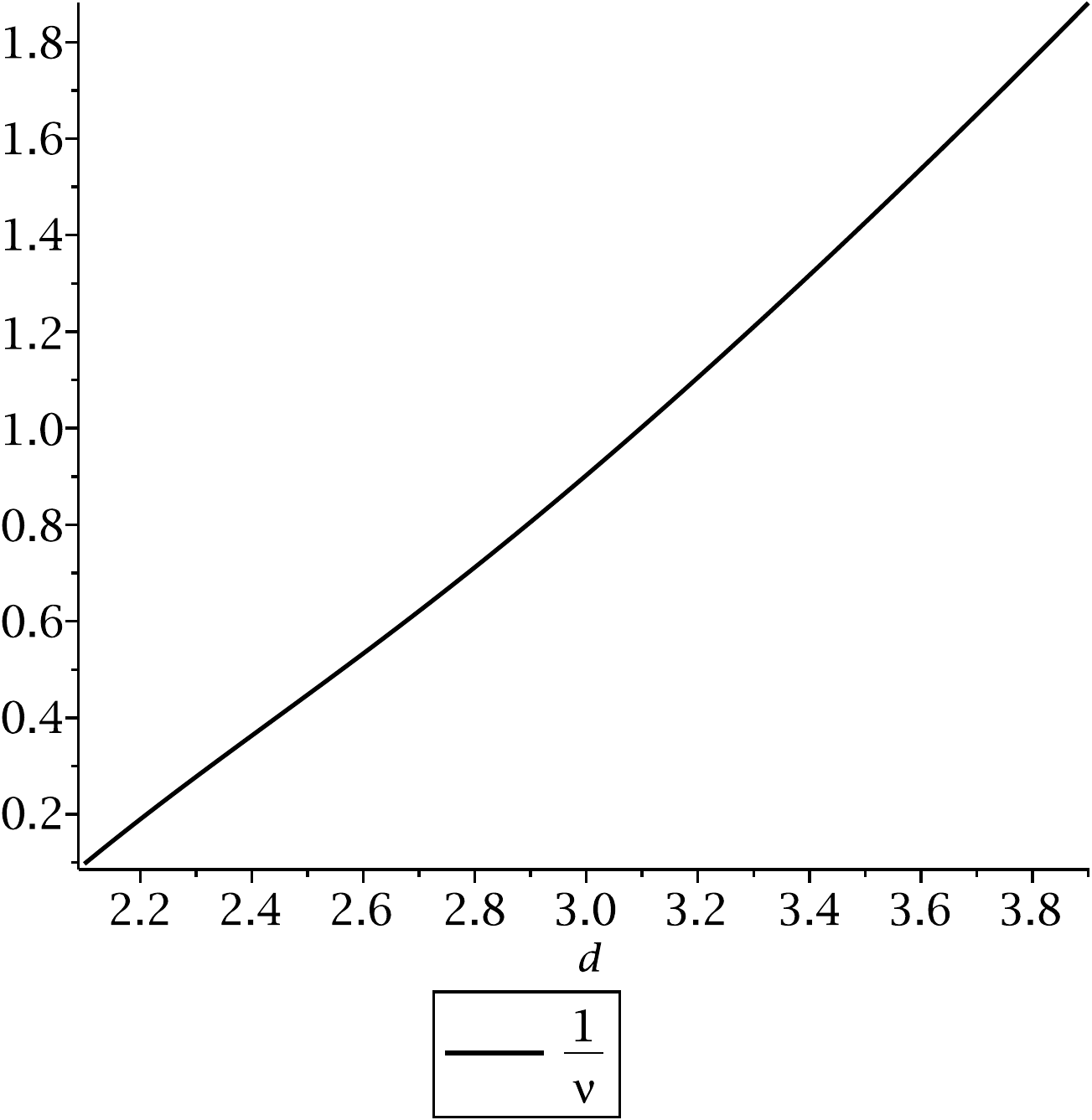}
\quad
\end{center}
\caption{Dependence of Pad\'{e} approximants to critical exponents $\eta$,
$\eta_\phi$ and $1/\nu$ for the Gross-Neveu XY model when $N$~$=$~$4$.}
\end{figure}}

\clearpage 

{\begin{figure}[hb]
\begin{center}
\includegraphics[width=7cm,height=7cm]{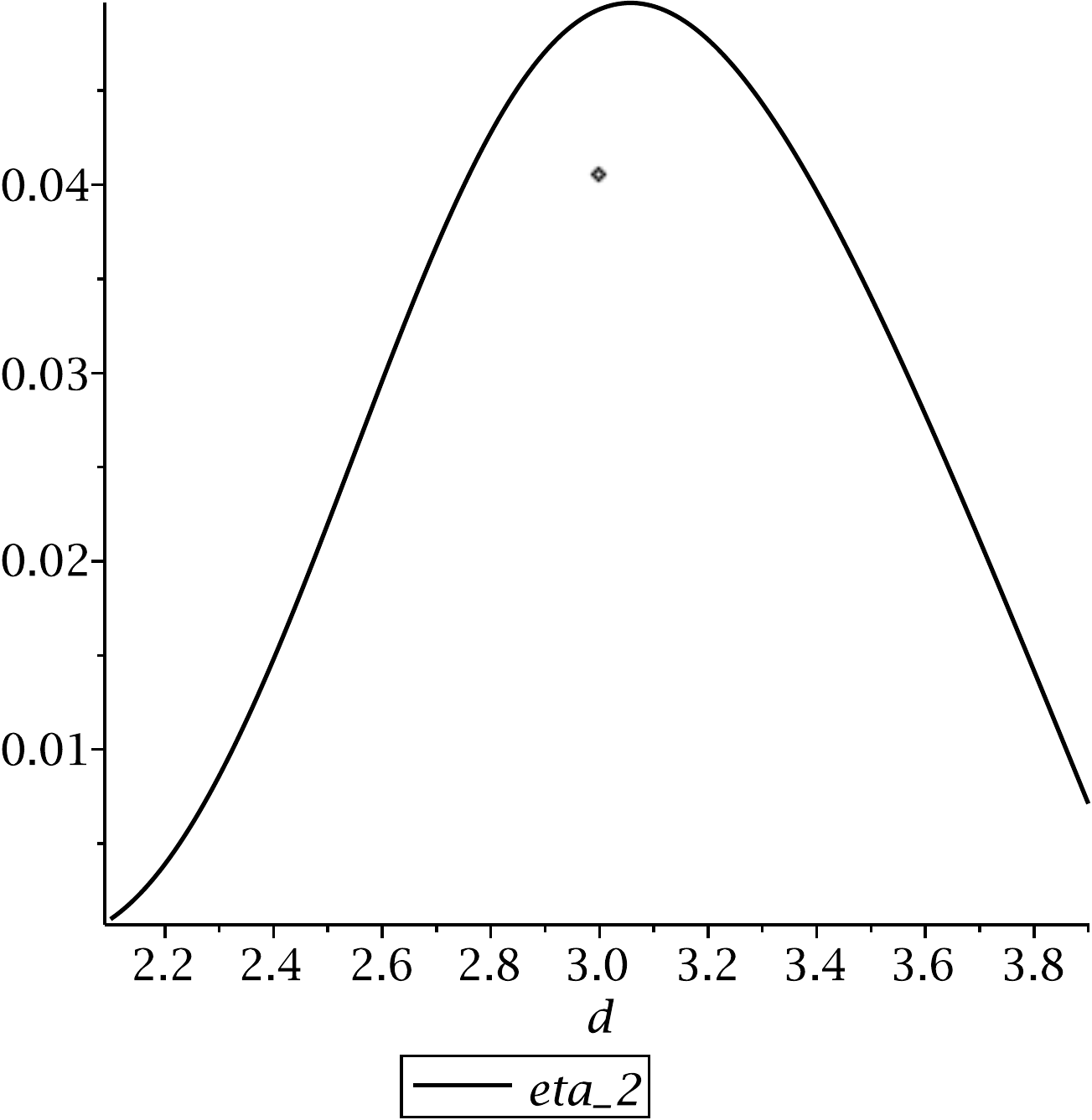}
\quad
\includegraphics[width=7cm,height=7cm]{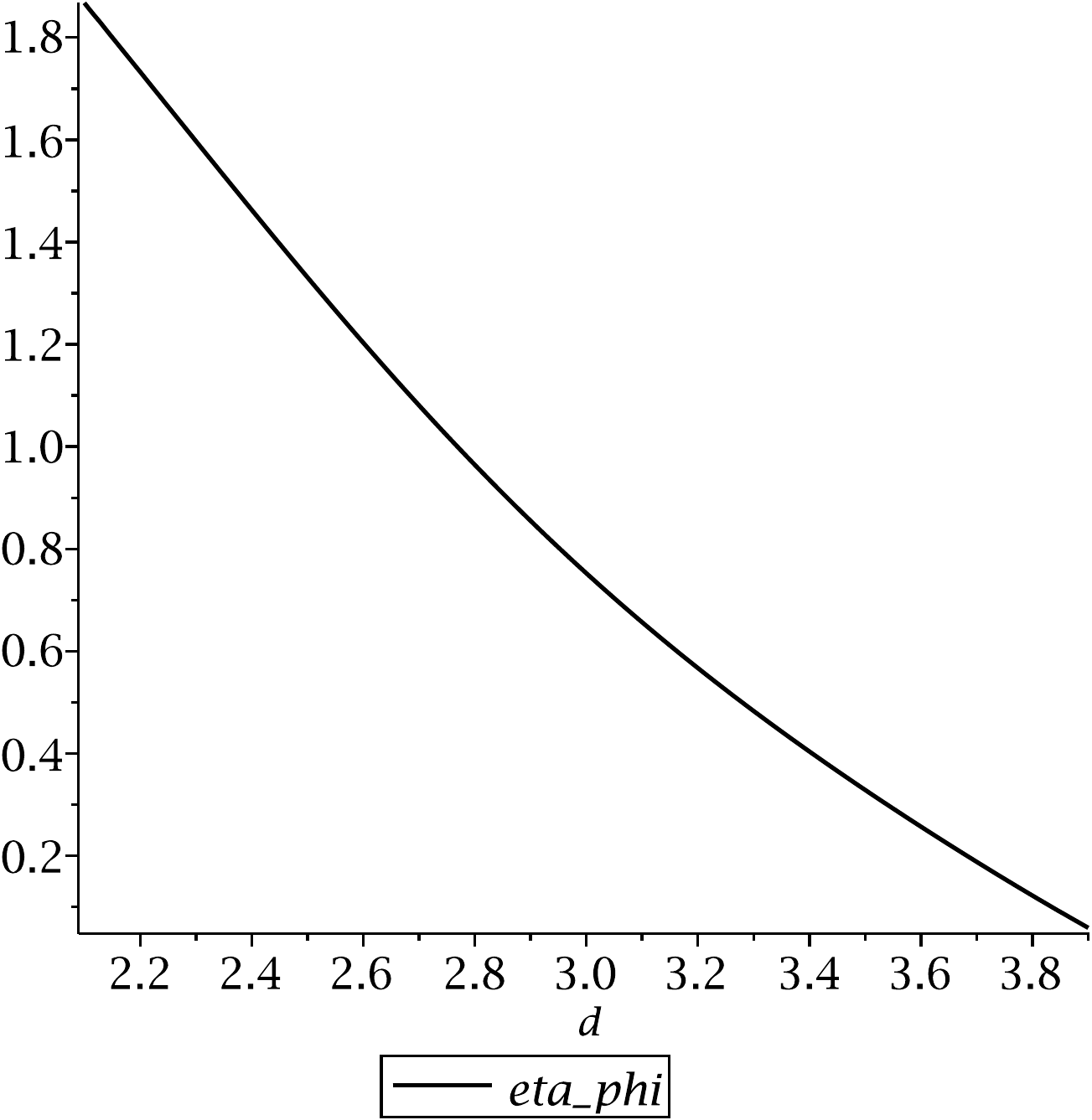} \\ 
\vspace{1.5cm}
\includegraphics[width=7cm,height=7cm]{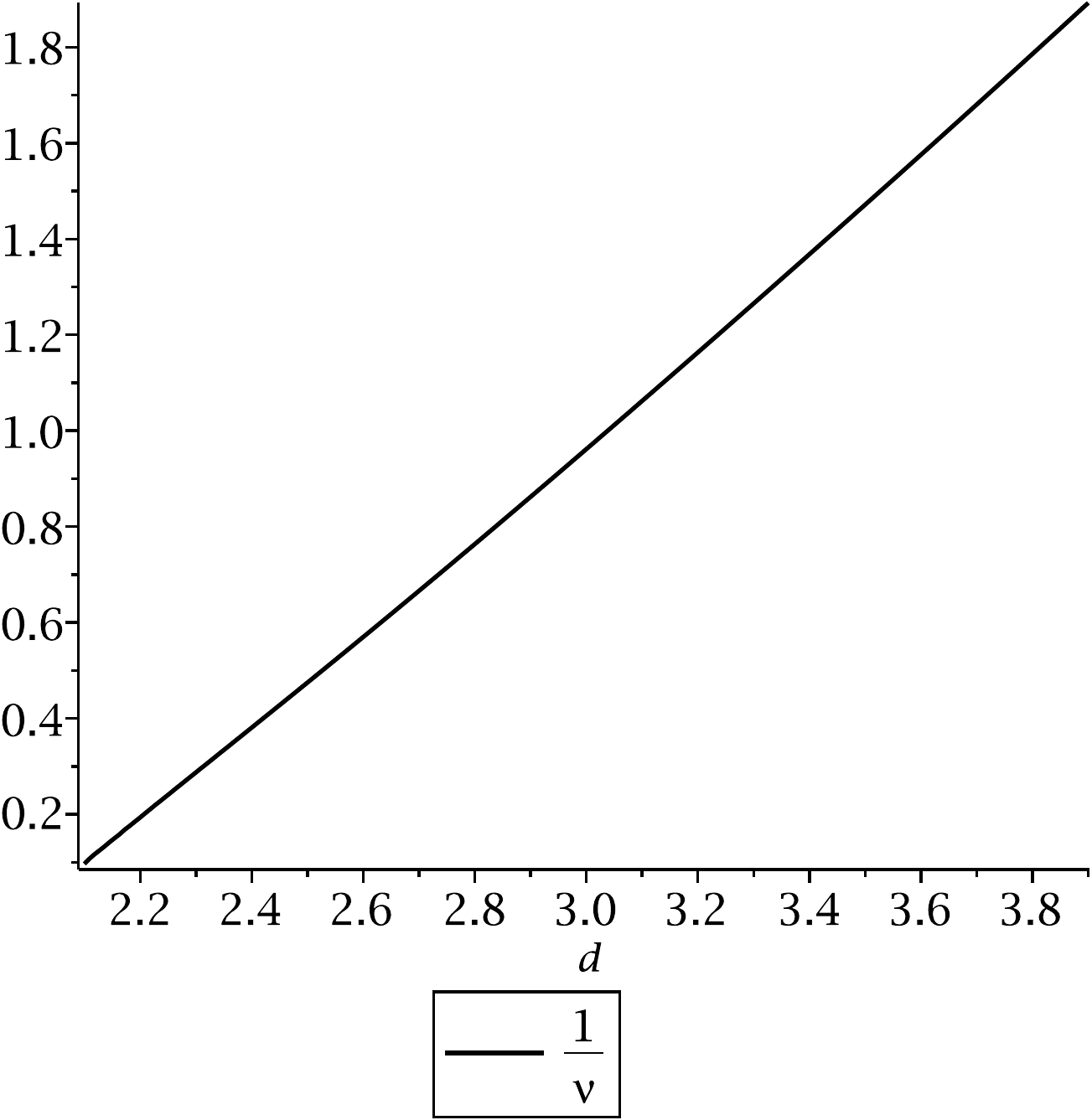}
\quad
\end{center}
\caption{Dependence of Pad\'{e} approximants to critical exponents $\eta$,
$\eta_\phi$ and $1/\nu$ for the Ising Gross-Neveu model when $N$~$=$~$8$.}
\end{figure}}

\sect{Discussion.}

We have completed the evaluation of the three core critical exponents $\eta$, 
$\eta_\phi$ and $1/\nu$ to several orders in the large $N$ expansion as a
function of the spacetime dimension $d$ for the chiral Heisenberg Gross-Neveu 
universality class. The $\epsilon$ expansion of the expressions near four
dimensions agrees exactly with the recent explicit four loop renormalization
group functions of (\ref{lagchgny}) at the Wilson-Fisher fixed point, given in
\cite{14}. Indeed our large $N$ results are an important independent check on 
that work. Moreover we have provided the next terms in the series as a future 
check for any five loop renormalization of (\ref{lagchgny}). As the exponents 
depend on $d$, plots of the Pad\'{e} approximant of each exponent in dimensions 
$2$~$<$~$d$~$<$~$4$ have been given in order to compare with the functional 
renormalization group approach of \cite{9} where a sharp regulator was used. 
All the large $N$ plots for not only the chiral Heisenberg Gross-Neveu 
universality class but the Ising and XY Gross-Neveu classes in $d$-dimensions 
are in good qualitative agreement with \cite{9}. Again this consistency 
provides independent evidence that these methods are capturing the proper and 
general behaviour of the exponents of the universality class across the 
dimensions. At the outset we drew attention to Figures $1$, $2$ and $3$ of 
\cite{9} in relation to the status of results available for the Ising and 
chiral Heisenberg Gross-Neveu universality classes. In addition to the three 
and four loop results of \cite{12,13,14} the results here are now edging towards
the latter class having commensurate data with the former. What is lacking is
Monte Carlo results and higher order two dimensional perturbative 
renormalization group functions. The latter should be possible to obtain to 
four loops with the recent derivation of the $\beta$-function for the Ising 
Gross-Neveu model in two dimensions, \cite{15}. This is not as straightforward 
a computation as that for the four dimensional case, \cite{12,13,14}. In two 
dimensions quartic fermion self-interactions are not multiplicatively 
renormalizable when the Lagrangian is dimensionally regularized. Instead 
additional evanescent quartic interactions are generated and their presence 
means that one has to be careful in extracting the true renormalization group 
functions after the regularization is lifted. Aside from this technical issue 
it should be possible in the future to add this extra information to the 
analysis of the chiral Heisenberg universality class.

\vspace{1cm}
\noindent
{\bf Acknowledgements.} The author thanks Prof I.F. Herbut, Dr L. Mihaila, Dr 
M. Scherer, R.M. Simms and Prof S.J. Hands for valuable discussions. The work 
was carried out with the support in part of the STFC through the Consolidated 
Grant ST/L000431/1. The graphs were drawn with the {\sc Axodraw} package 
\cite{53}. Computations were carried out in part using the symbolic 
manipulation language {\sc Form}, \cite{54,55}.

\end{document}